\documentclass[12pt]{article}
\usepackage{graphicx,fullpage,epsfig}

\newcommand{\lsim}{\raisebox{-0.13cm}{~\shortstack{$<$ \\[-0.07cm] $\sim$}}~}
\newcommand{\gsim}{\raisebox{-0.13cm}{~\shortstack{$>$ \\[-0.07cm] $\sim$}}~}

\newcommand{\s}{\smallskip}
\newcommand{\nn}{\noindent}

\newcommand{\beq}{\begin{eqnarray}}
\newcommand{\eeq}{\end{eqnarray}}
\newcommand{\tb}{\tan\beta}

\begin{document}
\baselineskip=18pt

\thispagestyle{empty}
\begin{flushright}
LPT--Orsay--06--07\\
PTA/06--03\\
UH-511-1080-06\\
January 2006
\end{flushright}

\vspace{1cm}

\begin{center}

{\large\sc {\bf Updated Constraints on the Minimal Supergravity Model}}

\vspace{1cm}

{\sc A. Djouadi$^{1}$, M. Drees$^2$} and {\sc Jean--Loic Kneur$^3$}

\vspace*{0.5cm}

$^1$ {\it Laboratoire de Physique Th\'eorique, UMR8627--CNRS, \\ 
Universit\'e Paris--Sud, Bt. 210, F--91405 Orsay, France.} 

\vspace*{3mm}

$^2${\it Dept. of Physics and Astronomy, University Of Hawaii, Honolulu,
  HI96822, USA} \\ and \\ {\it
Physikalisches Institut, Universit\"at Bonn, Nussallee 12, 53115 Bonn,
  Germany}\footnote{Permanent address.}.

\vspace*{3mm}

$^3${\it Laboratoire de Physique Th\'eorique et Astroparticules, UMR5207--CNRS,
\\ Universit\'e de Montpellier II, F--34095 Montpellier Cedex 5, France.}

\end{center}

\vspace{1cm}
\begin{abstract}
  
  \nn We provide an up--to--date analysis of the parameter space of the
  minimal supergravity model (mSUGRA). Novel features include the new central
  value of the top quark mass, an improved calculation of the masses of the
  supersymmetric particles and the neutral Higgs bosons, constraints from $b
  \rightarrow s \ell^+ \ell^-$ decays, and a careful treatment of the most
  important experimental and theoretical uncertainties. In addition to the by
  now traditional plots of the allowed region in the $(m_0, m_{1/2})$ plane,
  we show allowed regions in the planes spanned by pairs of {\em physical}
  sparticle or Higgs boson masses. Moreover, we search for the minimal
  allowed masses of new particles for various sets of constraints. We find
  that in many cases the direct experimental limits from collider and Dark
  Matter searches can be saturated even in this minimal model, and even after
  including the by now quite restrictive constraint on the Dark Matter relic
  density.

\end{abstract}

\newpage
\setcounter{page}{1}
\section*{1. Introduction} 

The minimal supergravity model (mSUGRA) \cite{mSUGRA,nilles} remains the
most widely studied implementation of the minimal supersymmetric extension of
the Standard Model (MSSM). It shares the virtues of a stable gauge hierarchy
(for sparticle masses not much above a TeV) \cite{hierarchy}, possible Grand
Unification of all gauge interactions \cite{GaugeUni}, and a plausible Dark
Matter (DM) candidate \cite{olddm,dmrev} with all variants of the
MSSM\footnote{A good DM candidate only emerges if $R-$parity is conserved,
  which we assume throughout.}. Moreover, it manages to describe
phenomenologically acceptable spectra with only four parameters plus a sign:
\beq \label{paras}
m_0, \ m_{1/2}, \ A_0, \ \tb, \ {\rm sign}{\mu}.
\eeq
Here $m_0$ is the common soft supersymmetry breaking contribution to the
masses of all scalars, $m_{1/2}$ the common supersymmetry breaking gaugino
mass, and $A_0$ the common supersymmetry breaking trilinear scalar interaction
(with the corresponding Yukawa coupling factored out); these three soft
breaking parameters are taken at the scale $M_X$ of Grand Unification, which
we define as the scale where the properly normalized $SU(2)_L$ and $U(1)_Y$
gauge couplings meet. Finally, $\tb$ is the ratio of the vacuum expectation
values (vev's) of the two Higgs doublets at the weak scale, which we identify
with the geometric mean of the soft breaking stop masses, and $\mu$ is the
supersymmetric higgs(ino) mass parameter.

It should be admitted that the choice of parameters (\ref{paras}) is not
particularly natural from a theoretical point of view: why should the scalar
masses and trilinear $A$ parameters all be exactly the same exactly at scale
$M_X$? From the perspective of supergravity theory, universality would seem to
emerge more naturally at a scale closer to the (reduced) Planck mass, $M_P
\simeq 2.4 \cdot 10^{18}$ GeV, if at all. However, while the possible
unification of the gauge interactions makes a strong argument for a ``grand
desert'' between the sparticle mass (or weak) and Grand Unified scales,
physics at energy scales above $M_X$ remains very speculative. At least as a
first approximation it is therefore not unreasonable to impose our boundary
conditions at scale $M_X$. 

The ansatz (\ref{paras}) also has important virtues, in addition to its
simplicity. It allows a quite varied phenomenology without violating any known
constraints. In particular, the assumed flavor universality implies that
supersymmetric flavor changing neutral current (FCNC) effects occur only
radiatively, through renormalization group (RG) evolution. This keeps FCNC
manageable, although, as we will see, flavor changing $b \rightarrow s \gamma$
and $b \rightarrow s \ell^+ \ell^-$ decays do impose important constraints on
the parameter space. A very appealing feature of mSUGRA is that it implements
radiative breaking of the electroweak gauge symmetry \cite{radbreak}, i.e. the
RG evolution naturally drives the squared mass of one of the Higgs fields to
negative values, keeping all squared sfermion masses positive. This allows to
determine the absolute value of $\mu$ as function of the other parameters.

In spite of these successes, in recent years there has been a proliferation of
analyses extending mSUGRA. Some of these extensions \cite{guts} are based on
specific Grand Unified models, and thus have independent motivation from
theory. However, in many phenomenological analyses universality between
sfermion masses and/or the universality of soft breaking Higgs and sfermion
masses is relaxed \cite{nonuniv} without specific theoretical
motivation\footnote{Bucking this trend, there have also been a few recent
  studies where additional relations between the parameters in (\ref{paras})
  are imposed \cite{rela}.}. Indeed, there seems to be a perception that the
parameter space of the model is getting ``squeezed'' by ever tightening
constraints.

Much of this perception probably comes from the by now quite accurate
determination of the relic density of Dark Matter (DM) particles. At least in
the framework of standard cosmology with a more or less scale invariant
primordial spectrum of density perturbations, the analysis of large
cosmological structures allows to infer the present DM density; the mapping of
the microwave sky by WMAP plays an especially important role here \cite{wmap}.
This translates into a quite tight constraint on mSUGRA parameter space
\cite{others} under the standard assumption that all DM is formed by lightest
superparticles (LSPs), which were in thermal equilibrium after the last period
of entropy production\footnote{In standard cosmology this means the end of the
  inflationary epoch.}.

It should be emphasized that this tight constraint should not be cause for
alarm. After all, the determination of the DM density, if taken at face value,
constitutes a genuine signal of physics beyond the Standard Model
(SM). Conceptually it should thus be considered on a par with, say, the
measurement of a selectron mass. The fact that mSUGRA can accommodate this
measurement is a further success of this model. 

Nevertheless it seems timely to re--assess the mSUGRA model, taking recent
theoretical and experimental developments into account. Besides the WMAP (and
related) cosmological data, these include:
\begin{itemize}
\item More accurate calculations of leading two--loop corrections to the
  masses of neutral Higgs bosons \cite{newhiggs}, which makes it somewhat
  easier to satisfy the stringent Higgs search limits from LEP for a fixed 
  value of the top quark mass;
\item The new, somewhat lower central value of the mass of the top quark
      \cite{newtop}, which in turn goes in the direction of decreasing the 
      predicted mass of the lightest neutral Higgs boson;
\item Improved limits on radiative $b$ decays and, in particular, first
  information on $b \rightarrow s \ell^+ \ell^-$ decays, which excludes
  scenarios where the sign of the amplitude of $b \rightarrow s \gamma$ decays
  is opposite to the SM prediction \cite{bsll};
\item A growing (though not global) consensus \cite{gmutalk} that the SM
  prediction for hadronic contributions to the anomalous dipole moment of the
  muon based on data from $e^+e^-$ colliders is more reliable, which again
  elevates the discrepancy between the measurement \cite{gmuexp} of $g_\mu -
  2$ and its SM prediction \cite{gmuth} to the level of $\sim 2.5$ standard
  deviations.
\end{itemize}

Several analyses of this kind have appeared in the last few years
\cite{others}, whose results broadly agree with our's if we take the old value
of the top mass, $m_t=178$ GeV. The effect of the new, reduced (central) value
of $m_t$ has so far only been analyzed in refs.~\cite{recent,nano-higgs}; these
papers have little overlap with our's. Usually the results of mSUGRA parameter
scans are presented as allowed regions in the $(m_0, m_{1/2})$ plane. We also
present similar allowed regions in the planes spanned by {\em physical}
sparticle or Higgs boson masses; this should give a more direct overview of
the kind of spectra that can be generated in mSUGRA. To the best of our
knowledge, similar results have previously only been published in the (by now
quite dated) review article \cite{dreesmartin}.

Moreover, we put special emphasis on a careful treatment of theoretical and
experimental uncertainties. This allows us to derive conservative lower
bounds on the masses of superparticles and Higgs bosons in mSUGRA, for
different sets of assumptions. We find that in many cases the direct
experimental search limits can be saturated even if all relevant constraints
are taken at face value. The main exceptions are the masses of the gluino and
of first and second generation squarks, which are forced by the assumption of
gaugino mass unification to lie at least 100 to 150 GeV above the current
Tevatron limits.  Imposing the DM constraint does {\em not} affect these lower
bounds very much.

The remainder of this article is organized as follows. In the next chapter we
briefly describe calculational details and the constraints we impose. Sec.~3
updates ref.~\cite{ddk1} by showing the allowed regions in the $(m_0,
m_{1/2})$ plane, as well as in the planes spanned by pairs of physical masses,
for a few values of $\tb$. Sec.~4 is devoted to a discussion of minimal masses
of sparticles and Higgs bosons that are compatible with various sets of
constraints. Finally, in Sec.~5 we summarize our main results and draw some
conclusions.

\section*{2. Scanning Procedures}

We use the FORTRAN program SuSpect \cite{suspect} to calculate the spectrum of
superparticles and Higgs bosons. Since these masses are defined at the weak
scale while the dimensionful input parameters in (\ref{paras}) are defined at
the scale of Grand Unification, the program has to integrate the set of
coupled renormalization group equations connecting these two scales; SuSpect
now uses two--loop equations \cite{rge2l} for all relevant quantities (gauge
and Yukawa couplings, $\mu$, and the soft breaking parameters). The program
also computes the one--loop and dominant two--loop corrections to the Higgs
potential, as well as the dominant one--loop corrections turning the running
($\overline{\rm DR}$) quark, lepton and sparticle masses into on--shell (pole)
masses. The calculation of the neutral Higgs boson masses includes leading
two--loop corrections \cite{newhiggs}.  See ref.~\cite{suspect} for further
details on the calculation of the spectrum.\s

Not all combinations of input parameters lead to radiative $SU(2)_L \times
U(1)_Y \rightarrow U(1)_{\rm QED}$ symmetry breaking. This imposes a first
constraint on the parameter space. We also exclude parameter sets where the
scalar potential has deep minima breaking charge and/or color {\em at the weak
  scale} \cite{oldccb}. As usual in the literature \cite{others}, we do not
veto scenarios where the absolute minimum of the scalar potential occurs for
field values intermediate between the weak and GUT scales \cite{casas}, since
the tunneling rate into these minima is exceedingly slow. In the language of
ref.~\cite{casas}, we impose the ``CCB'' constraints, which exclude very large
values of $|A_0|/\sqrt{m_0^2 + m_{1/2}^2}$, but do not impose the ``UFB''
constraints.

We next impose experimental constraints. To begin with, the strong upper
limits on the abundance of stable charged particles \cite{pdg} exclude
scenarios where the lightest superparticle is charged. This excludes cases
with $m_0 \ll m_{1/2}$ where $\tilde \tau_1$ tends to be the LSP (especially
at large $\tb$), and some combinations with $m_0 \gsim m_{1/2}$ and sizable
$|A_0|$, where $\tilde t_1$ is the LSP\footnote{These constraints can be
  evaded if the LSP resides in the hidden sector (e.g., if it is the gravitino
  \cite{gravi}), or in extensions of mSUGRA where the LSP is an axino
  \cite{axi}.}.

We also impose the lower bounds on sparticle and Higgs masses that result from
collider searches. We interpret the LEP limits from searches for (unstable)
charged superparticles as requiring
\beq \label{xseclim}
\sigma(e^+ e^- \rightarrow \tilde X \bar{\tilde X} ; \sqrt{s} = 209 \ {\rm
  GeV}) < 20 \ {\rm fb}
\eeq
separately for each relevant mode ($\tilde X = \tilde t_1, \tilde \tau_1,
\tilde \chi_1^+$). This effectively imposes lower bounds of 104.5 GeV on the
mass of the lighter chargino $\tilde \chi_1^+$, 101.5 GeV for the lighter
scalar top eigenstate $\tilde t_1$, and 98.8 GeV for the lighter scalar $\tau$
eigenstate $\tilde \tau_1$. For non--pathological situations these agree
closely with the limits published by the LEP experiments
\cite{pdg,lepsusy}.\footnote{The only ``pathological'' situation that can be
  relevant in mSUGRA is the case of small $\tilde \tau_1 - \tilde \chi_1^0$
  mass splitting. However, at small $\tan\beta$ selectron searches at LEP will
  lead to constraints on the parameter space that are nearly as strong as
  those from $\tilde \tau_1$ searches. At high $\tan\beta$, scenarios with
  small slepton and neutralino masses are excluded by the $g_\mu$ constraint.} 

The limits from the searches for Higgs bosons at LEP also impose important
constraints on the parameter space. Of special importance is the limit on $e^+
e^- \rightarrow Z H$ with $H \rightarrow b \bar b$. In the SM it leads to the
bound \cite{lephiggs} $m_{H}^{\rm SM} > 114$ GeV. For small and intermediate
values of $\tb$ this bound applies directly to the light scalar Higgs boson in
mSUGRA, but for $\tb \gsim 50$ its coupling to the $Z$ boson can be suppressed
significantly. We parameterize this dependence as in ref.~\cite{ddk1}, except
that the constant (coupling--independent) term is increased by 0.5 GeV in
order to reflect the increase of the limit that resulted from combining the
limits from the four LEP experiments. 

We also include constraints from quantum corrections due to superparticles. 
These include the upper bound
\beq \label{delrho}
\delta \rho_{\rm SUSY} < 2.2 \cdot 10^{-3}
\eeq
on the supersymmetric contribution to the electroweak $\rho-$parameter
\cite{rho}, including 2--loop QCD corrections \cite{rho2}. However, it turns
out that this constraint is always superseded by either the LEP Higgs search
limit or by the CCB constraint.

A more significant constraint arises from the precise measurements of the
anomalous magnetic moments of positively and negatively charged muons
\cite{gmuexp}. As well known by now, the interpretation of this measurement
hinges on whether data from semileptonic $\tau$ decays are used for the
evaluation of the SM prediction or if one only relies on data from $e^+e^-$
annihilation into hadronic final states. In the former case the measurement
agrees quite well with the SM, whereas in the latter case the prediction falls
$\sim 2.5 \sigma$ short of the measurement \cite{gmuth}. In order to reflect
this uncertainty, we impose either the more conservative constraint
\beq \label{gmuc1}
-5.7 \cdot 10^{-10} \leq a_{\mu,\, {\rm SUSY}} \leq 4.7 \cdot 10^{-9},
\eeq
which describes the overlap of the $2\sigma$ limits derived from the two
competing SM predictions, or the more aggressive constraint
\beq \label{gmuc2}
1.06 \cdot 10^{-9} \leq a_{\mu,\, {\rm SUSY}} \leq 4.36 \cdot 10^{-9},
\eeq
which is the 90\% c.l. range derived using the $e^+e^-$ data only. Here
$a_{\mu,\,{\rm SUSY}}$ is the sparticle loop contribution to $a_\mu \equiv
(g_\mu - 2)/2$. The SM prediction based on $e^+e^-$ data is nowadays
considered to be more reliable \cite{gmutalk}. Note that (\ref{gmuc1}) allows
the supersymmetric contribution to vanish, or even be slightly negative,
whereas (\ref{gmuc2}) requires it to be positive. Our calculation of
$a_{\mu,\,{\rm SUSY}}$ is based on ref.~\cite{gmususy}, modified to include
leading logarithmic QED 2--loop corrections \cite{pepe2} which increase
$a_{\mu,\, {\rm SUSY}}$ by $\sim 5\%$.

The constraints discussed so far are all quite robust against minor changes of
the model. In particular, deviating from exact universality of scalar masses
or, equivalently, allowing small flavor non--diagonal entries of the sfermion
mass matrices, will not change any of these bounds significantly. This is
quite different for the bounds from inclusive $b \rightarrow s \gamma$ decays,
which are also widely included in analyses of the parameter space of mSUGRA
and similar models \cite{guts, nonuniv, rela, others}. Including theoretical
uncertainties of the SM prediction \cite{bsgsm} as well as the experimental
measurement \cite{pdg} (now statistically dominated by BELLE data), we require
the calculated branching ratio to fall in the range
\beq \label{bsg}
2.65 \cdot 10^{-4} \leq B(b \rightarrow s \gamma) \leq 4.45 \cdot 10^{-4}.
\eeq
Our calculation of this branching ratio is based on ref.~\cite{gambino}, which
-- for heavy sparticles -- includes the dominant QCD corrections to the
$\tilde \chi^\pm \tilde t$ loop corrections, which (together with $t H^\pm$
loops) dominate the supersymmetric contributions in all supersymmetric models
where flavor violation is assumed to be described entirely by the
Kobayashi--Maskawa (KM) matrix. For large $\tan\beta$ and not too heavy
sparticles these contributions can be quite large; for $\mu > 0$ they can even
flip the sign of the amplitude relative to the SM prediction, leading to a
second allowed region \cite{ddk1} when the modulus of this amplitude
approaches its SM value. However, it has recently been argued \cite{bsll} that
new data on $b \rightarrow s \ell^+ \ell^-$ data strongly disfavor this
possibility, since such a flip of sign would change the interference between
penguin diagrams (similar to those contributing to $b \rightarrow s \gamma$)
and (new) box diagrams. We therefore impose the additional constraint that the
amplitude for $b \rightarrow s \gamma$ decays should have the sign predicted
in the SM.

The range (\ref{bsg}) should perhaps be extended somewhat, since the MSSM
prediction has larger theoretical uncertainties than that in the SM. To begin
with, ref.~\cite{gambino} includes NLO QCD corrections to the supersymmetric
contribution only in the limit of heavy sparticles, as remarked above. Note
also that the determination of the KM element $V_{ts}$, to which all
contributions to the amplitude describing $b \rightarrow s \gamma$ decays are
proportional, can be affected by supersymmetric contributions to processes in
the $K$ sector, which in the SM lead to tight constraints on this quantity.
More importantly, the constraint imposed by (\ref{bsg}) on the parameters
listed in (\ref{paras}) will evaporate entirely if a modest amount of $\tilde
b - \tilde s$ mixing is allowed at the GUT scale \cite{leszek}. This would
lead to one--loop gluino--mediated box diagram contributions to $b \rightarrow
s \gamma$ decays \cite{bsggl}. Since for strict scalar universality all
contributions are suppressed by a factor $|V_{ts}| \simeq 0.04$, even a small
amount of $\tilde b - \tilde s$ mixing would lead to new contributions of
comparable size. The sign of this new contribution is given by the sign of the
corresponding mixing angle, which is a free parameter in this slightly
extended model. At the price of modest fine--tuning one could thus make any
set of mSUGRA parameters ``$b \rightarrow s \gamma$ compatible''. Since the
required squark flavor mixing would still be quite small, it would have
negligible effects on (flavor conserving) signals at colliders, radiative
corrections to the MSSM Higgs sector, etc. The constraint (\ref{bsg}) thus has
a different status from the constraints discussed earlier. In Sec.~4 we will
therefore present limits on sparticle and Higgs boson masses with or without
this additional constraint.

The last, and very restrictive, constraint that is usually imposed in analyses
of constrained supersymmetric models is based on the determination of the
density of non--baryonic Dark Matter (DM) from detailed analyses of the
anisotropies of the cosmic microwave background (CMB), in particular by the
WMAP experiment \cite{wmap}. At 99\% c.l., they find
\beq \label{dm}
0.087 \leq \Omega_{\rm DM} h^2 \leq 0.138.
\eeq
Here $\Omega$ measures the mass (or energy) density in units of the critical
density, whereas $h$ is the scaled Hubble constant. The emergence of the
cosmological ``concordance model'' is undoubtedly a great triumph of modern
cosmology. One should nevertheless be aware that the result (\ref{dm}) is
based on several assumptions, which are reasonable but not easy to
cross--check independently \cite{susytalk}. In particular, one has to assume
that simple inflationary models give essentially the right spectrum of
primordial density perturbations. Without this, or a similarly restrictive,
ansatz for this spectrum, the result (\ref{dm}) would evaporate. In the
absence of a generally accepted estimate of the theoretical uncertainty from
the assumed ansatz of the primordial density perturbations, we decided to take
the 99\% c.l. region of the DM relic density, as opposed to the 90\% or 95\%
c.l. regions used for quantities measured in the laboratory.

In order to translate the constraint (\ref{dm}) into a constraint on mSUGRA
parameters, one has to make several {\em additional} assumptions. The lightest
neutralino $\tilde \chi_1^0$ must be essentially stable, which, in the context
of mSUGRA with conserved $R-$parity, requires it to be heavier than the
gravitino \cite{gravi}. In addition, one must assume that standard cosmology
(with known Hubble expansion rate, and no epoch of entropy production) can be
extrapolated backwards to temperatures of at least $\sim 5\%$ of $m_{\tilde
  \chi_1^0}$. In that case $\tilde \chi_1^0$ was in full thermal equilibrium
with the SM plasma, making today's $\tilde \chi_1^0$ density independent of
the ``re--heat'' temperature of the Universe $T_R$ at the end of inflation.
With these assumptions, our calculation of the relic density proceeds as in
ref.~\cite{ddk1}. Without these assumptions, no meaningful constraint on the
parameters (\ref{paras}) results. In Sec.~4 we therefore again present results
with or without the constraint (\ref{dm}).

\section*{3. Results of the scans}

\subsection*{3.1 ($m_0, m_{1/2}$) parameter space}

Examples of scans of the $(m_0, m_{1/2})$ plane are shown in Figs.~1 through
6. Figs.~1 and 2 represent our ``base choice'', $m_t = 172.7$ GeV (the current
central value \cite{newtop}), $A_0 = 0$ and $\mu > 0$. The light grey regions
are excluded by theoretical constraints (in particular, by the requirement of
correct electroweak symmetry breaking), as well as by the searches for
sparticles, i.e. by the constraint (\ref{xseclim}). The dark grey areas are
excluded by the requirement that the LSP be neutral, in particular by
$m_{\tilde\tau_1} > m_{\tilde\chi_1^0}$. The pink regions are excluded by
searches for neutral Higgs bosons at LEP. The light pink regions are excluded
even if we allow for a 3 GeV theoretical uncertainty in the calculation of
$m_h$ in the MSSM, i.e. if we only exclude (SM--like) Higgs bosons with
calculated mass $m_h \leq 111$ GeV. In the medium pink region the predicted
$m_h$ lies between 111 and 114 GeV, which might be acceptable if unknown
higher order corrections are sufficiently large and positive. As well known,
the LEP data \cite{lephiggs} show some (weak) evidence for the existence of an
SM--like Higgs boson with mass near 114 GeV; the mSUGRA regions that can
explain this small excess of Higgs--like events are shown in red.

\begin{figure}[h!]
\hspace*{.5cm} ${\mathbf m_0}$ ({\bf GeV})
\begin{center}
\includegraphics[width=.83\textwidth]{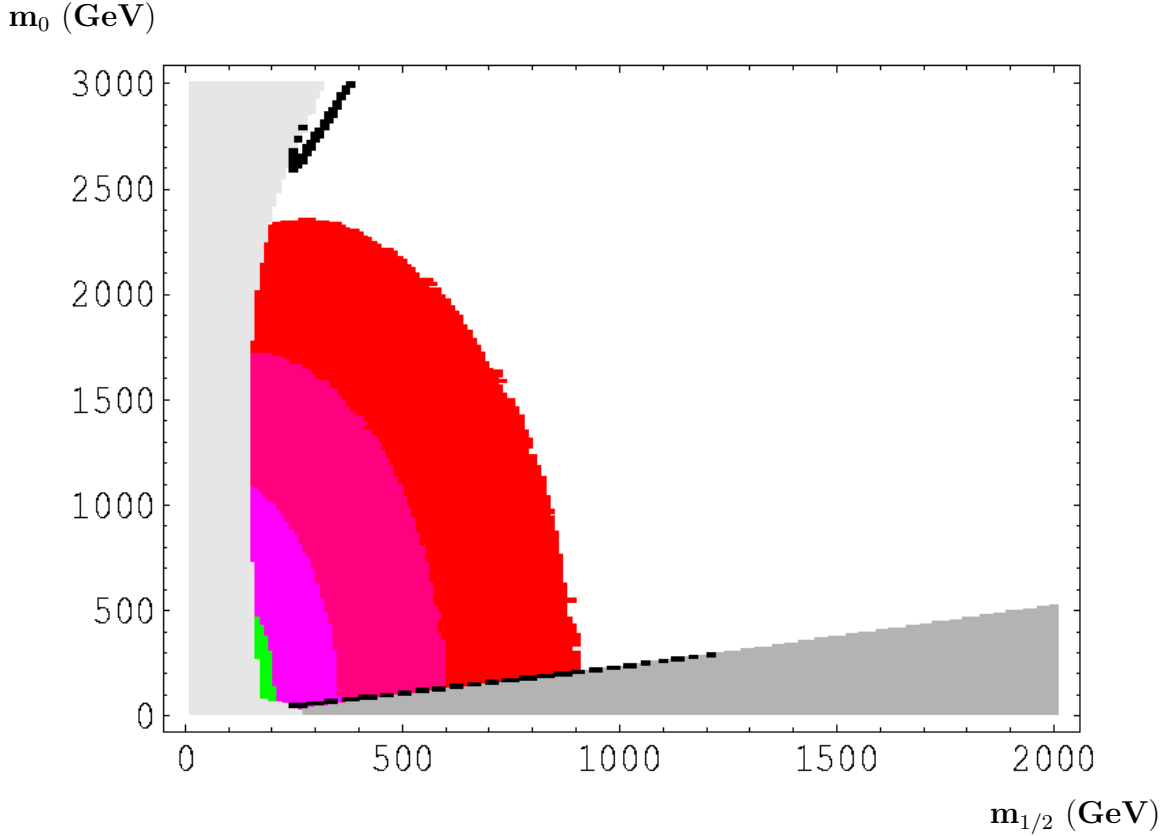}
\hspace*{13cm} ${\mathbf m_{1/2}}$ ({\bf GeV}) 
\caption{\it The mSUGRA $(m_{1/2},m_0)$ parameter space with all constraints 
  imposed for $A_0=0, \mu>0$ and $\tan\beta=10$. The top quark mass is fixed
  to the new central value, $m_t=172.7$ GeV. The light grey region is
  excluded by the requirement of correct electroweak symmetry breaking, or by
  sparticle search limits. In the dark grey region $\tilde\tau_1$ would be
  the LSP. The light pink region is excluded by searches for neutral Higgs
  bosons at LEP, whereas the green region is excluded by the $b \rightarrow
  s \gamma$ constraint (\ref{bsg}). In the blue region, the SUSY contribution
  to $g_\mu-2$ falls in the range (\ref{gmuc2}), whereas the red regions are
  compatible with having an SM--like Higgs boson near 115 GeV. Finally, the
  black regions satisfy the DM constraint (\ref{dm}).}
\end{center}
\vspace*{-.5cm}
\end{figure}

\begin{figure}[h!]
${\mathbf m_0}$
\vspace*{-.3cm}
\begin{center}
\mbox{\epsfig{file=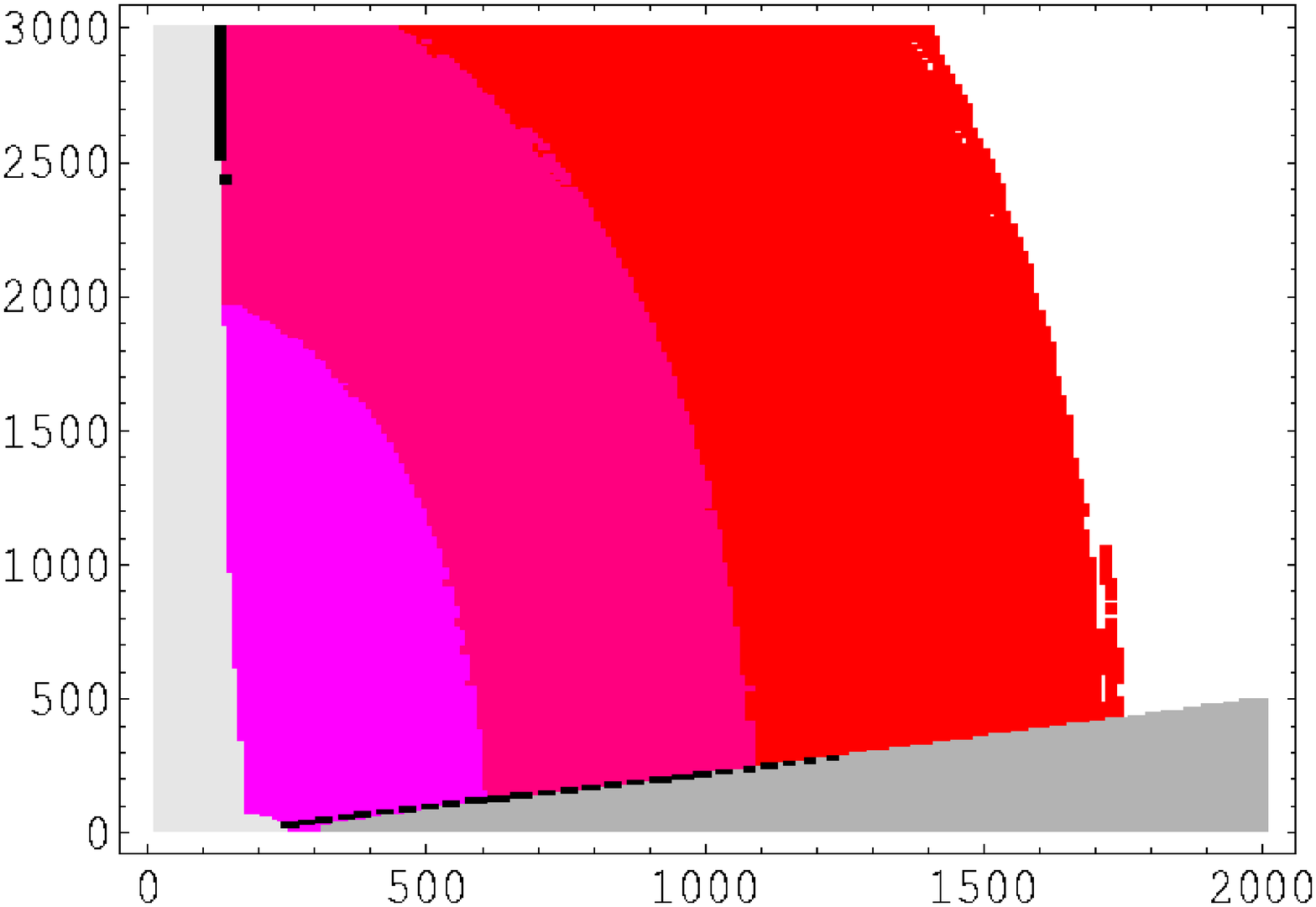,width=7.5cm,height=7.cm}\hspace*{0.9cm}
      \epsfig{file=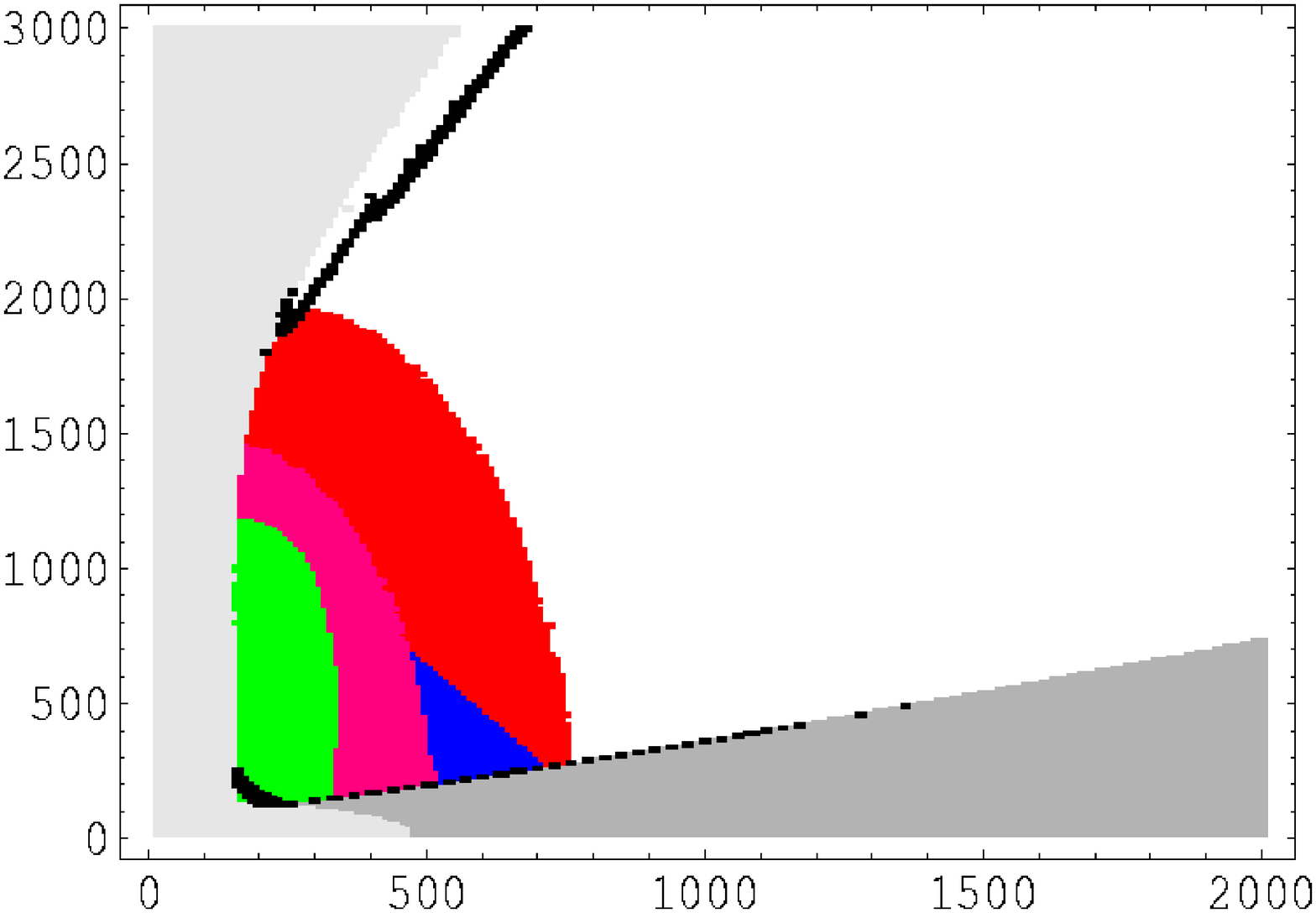,width=7.5cm,height=7.cm} }
\hspace*{13.9cm} ${\mathbf m_{1/2}}$
\vspace*{-.3cm}
\end{center}
${\mathbf m_0}$
\vspace*{-.3cm}
\begin{center}
\mbox{\epsfig{file=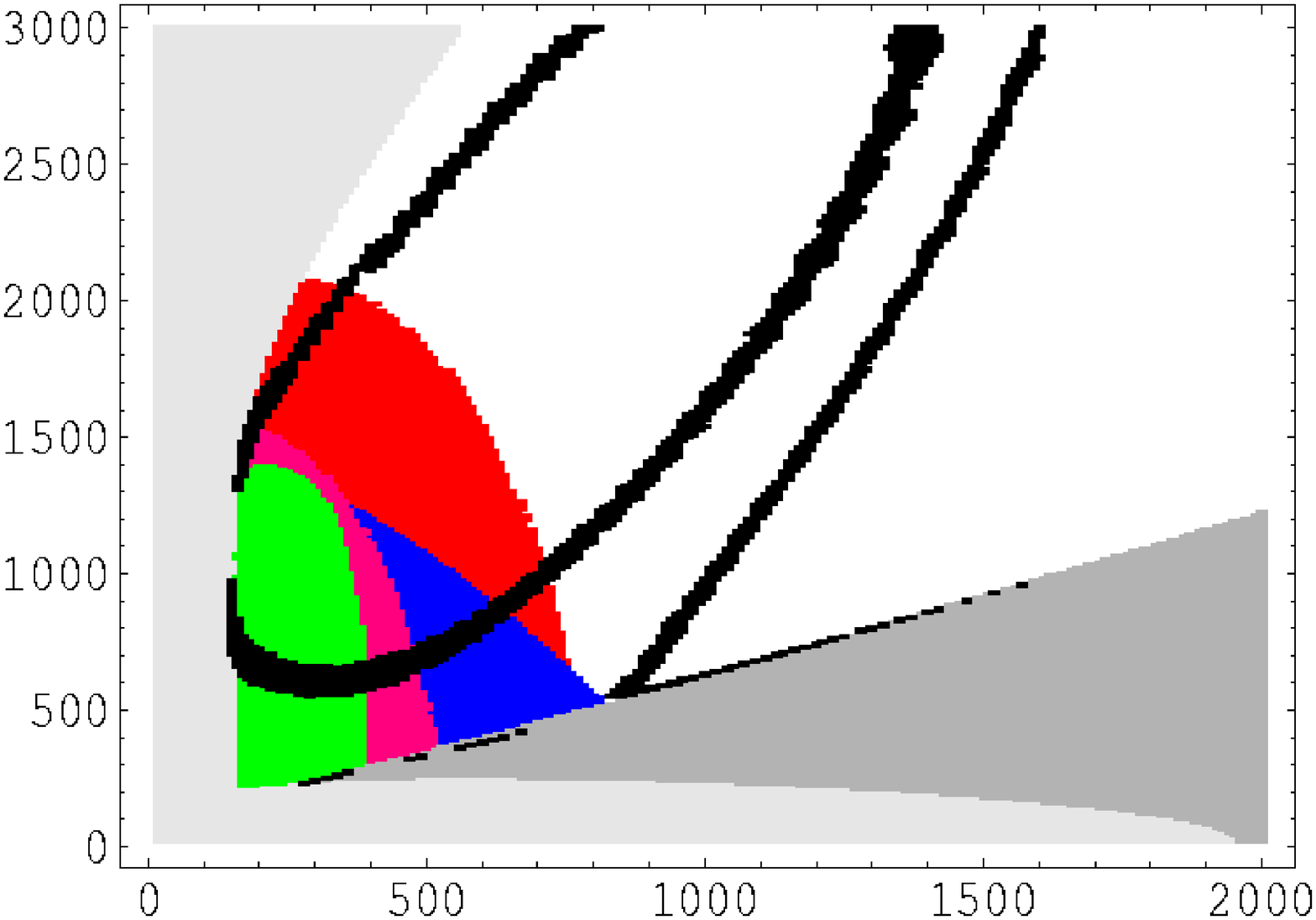,width=7.5cm,height=7.cm}\hspace*{0.9cm}
      \epsfig{file=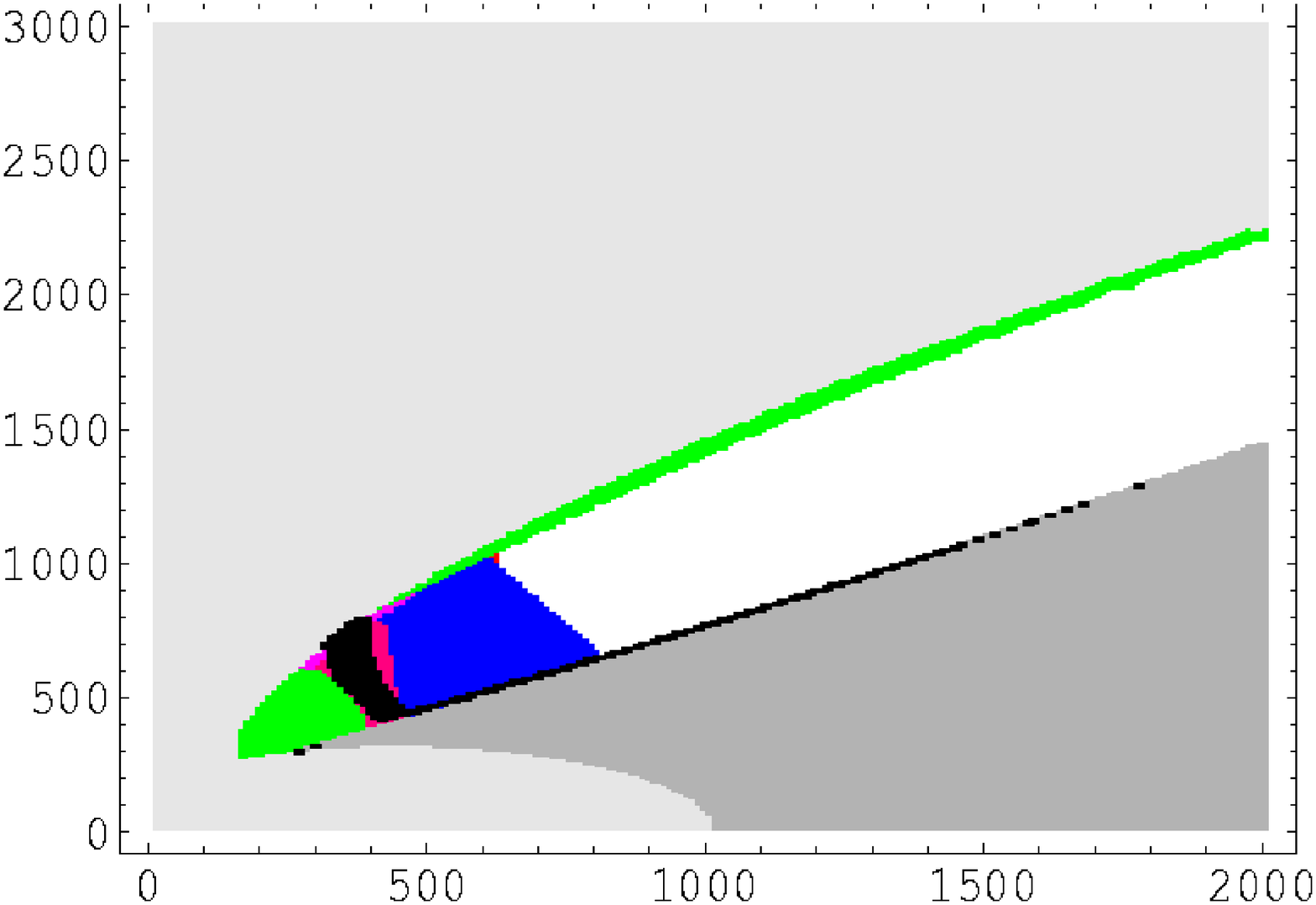,width=7.5cm,height=7.cm} }
\hspace*{13.9cm} ${\mathbf m_{1/2}}$
\end{center}
\vspace*{-.3cm}
\caption{\it The mSUGRA $(m_{1/2},m_0)$ parameter space with all constraints 
  imposed for $A_0=0, \mu>0$ and the values $\tan\beta=5$ (top left), 30 (top
  right), 50 (bottom left) and 58 (bottom right). The top quark mass is fixed
  to the new central value, $m_t=172.7$ GeV. Notation and
  conventions are as in Fig.~1.}
\end{figure}

In the blue regions the SUSY contribution to the anomalous magnetic moment of
the muon falls in the range (\ref{gmuc2}) favored by $e^+e^-$ data; recall
that in this case a positive SUSY contribution is required at the $\sim 2.5 \,
\sigma$ level. The green region is excluded by the constraint (\ref{bsg}) on
the branching ratio for radiative $b \rightarrow s \gamma$ decays. Finally, in
the black regions the $\tilde \chi_1^0$ relic density lies in the desired
range (\ref{dm}).

\begin{figure}[h!]
${\mathbf m_0}$
\vspace*{-.3cm}
\begin{center}
\mbox{\epsfig{file=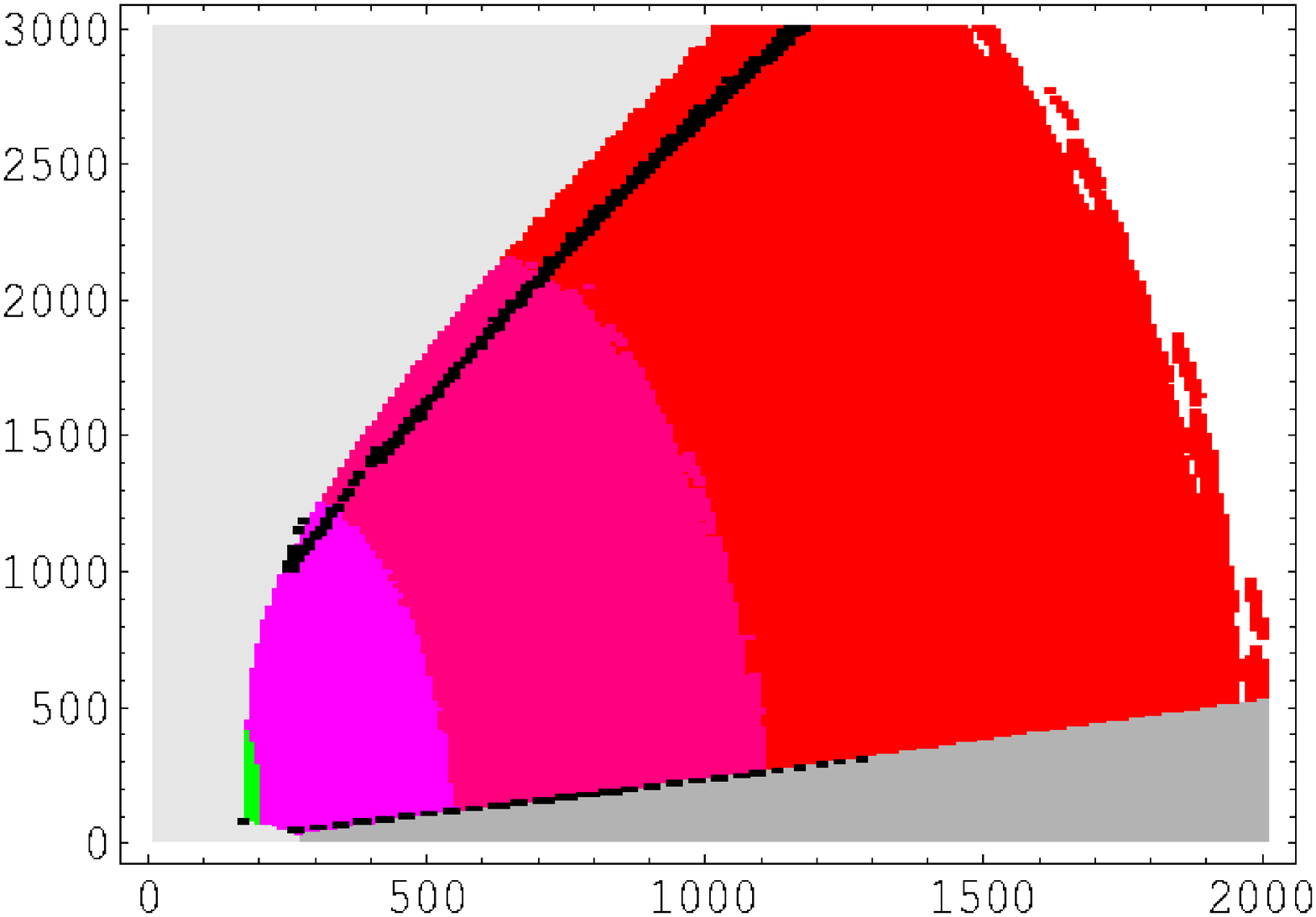,width=7.5cm,height=7.5cm}\hspace*{0.9cm}
      \epsfig{file=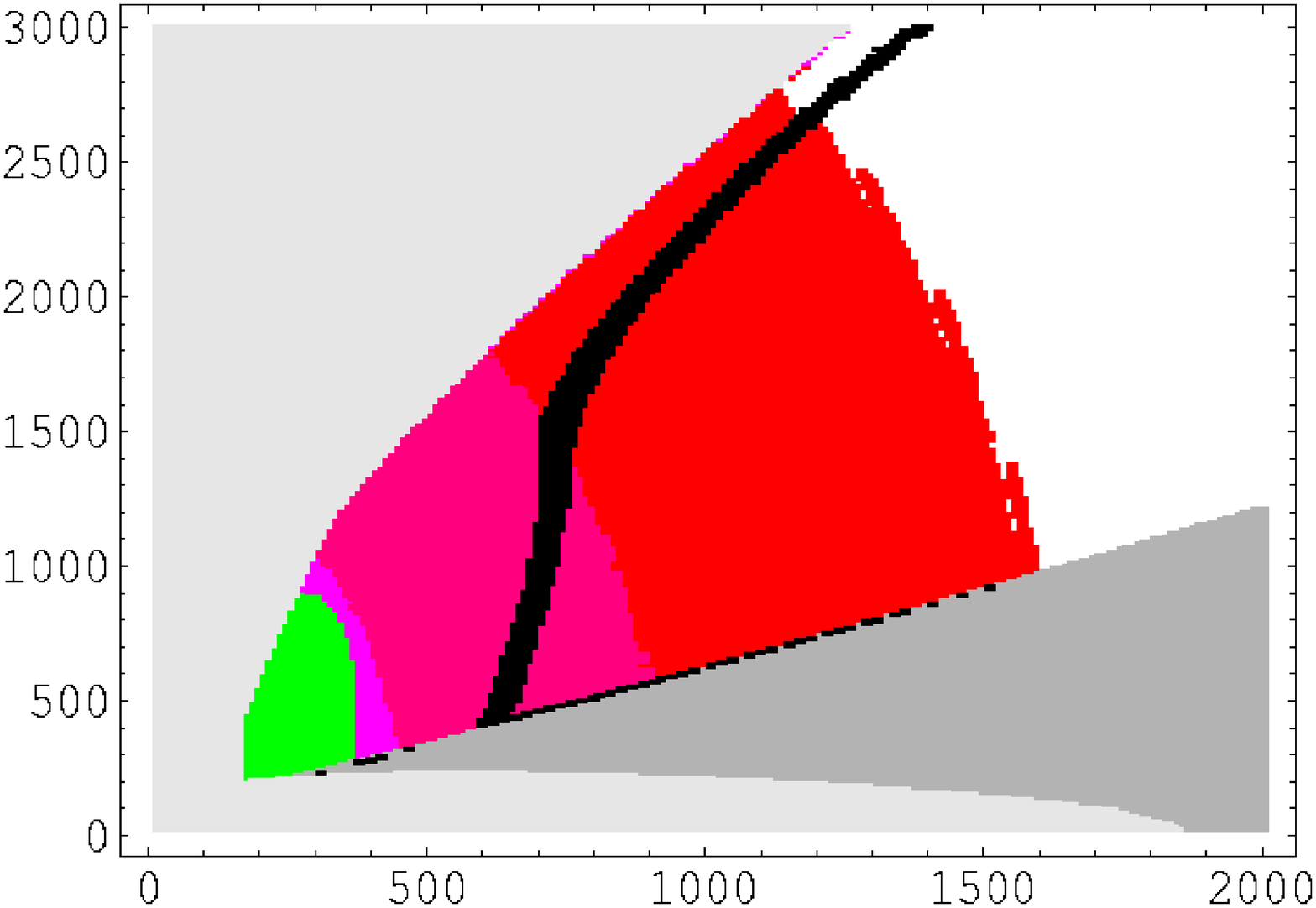,width=7.5cm,height=7.5cm} }
\hspace*{13.9cm} ${\mathbf m_{1/2}}$
\vspace*{-.3cm}
\end{center}
\caption{\it The mSUGRA $(m_{1/2},m_0)$ parameter space with all constraints 
  imposed for $A_0=0, \mu>0$ and the values $\tan\beta=10$ (left) and 50
  (right). The top quark mass is fixed to $m_t=166.9$ GeV. Notation and
  conventions are as in Fig.~1.}
\vspace*{.3cm}
\end{figure}

\begin{figure}[h!]
${\mathbf m_0}$
\vspace*{-.3cm}
\begin{center}
\mbox{\epsfig{file=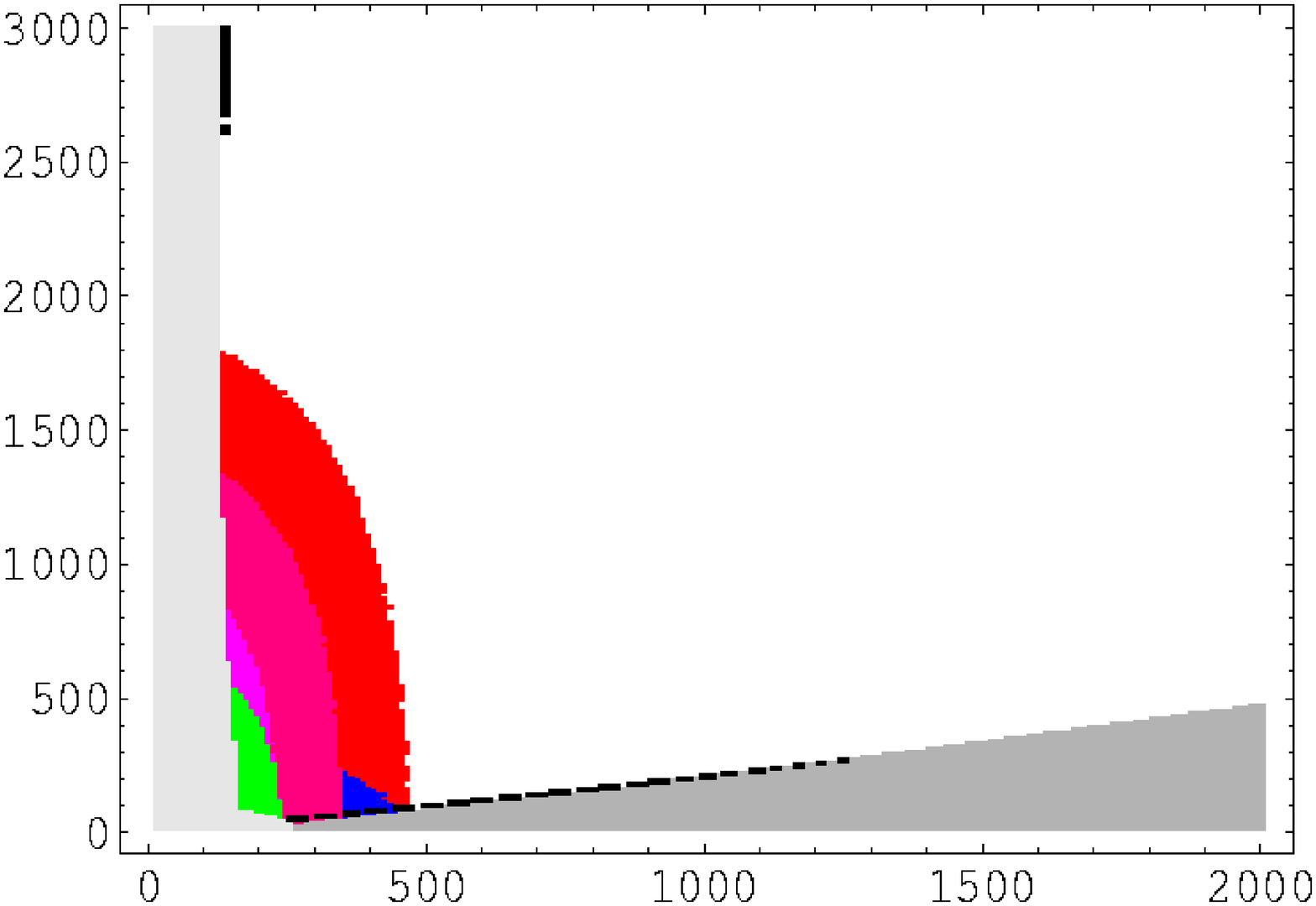,width=7.5cm,height=7.5cm}\hspace*{0.9cm}
      \epsfig{file=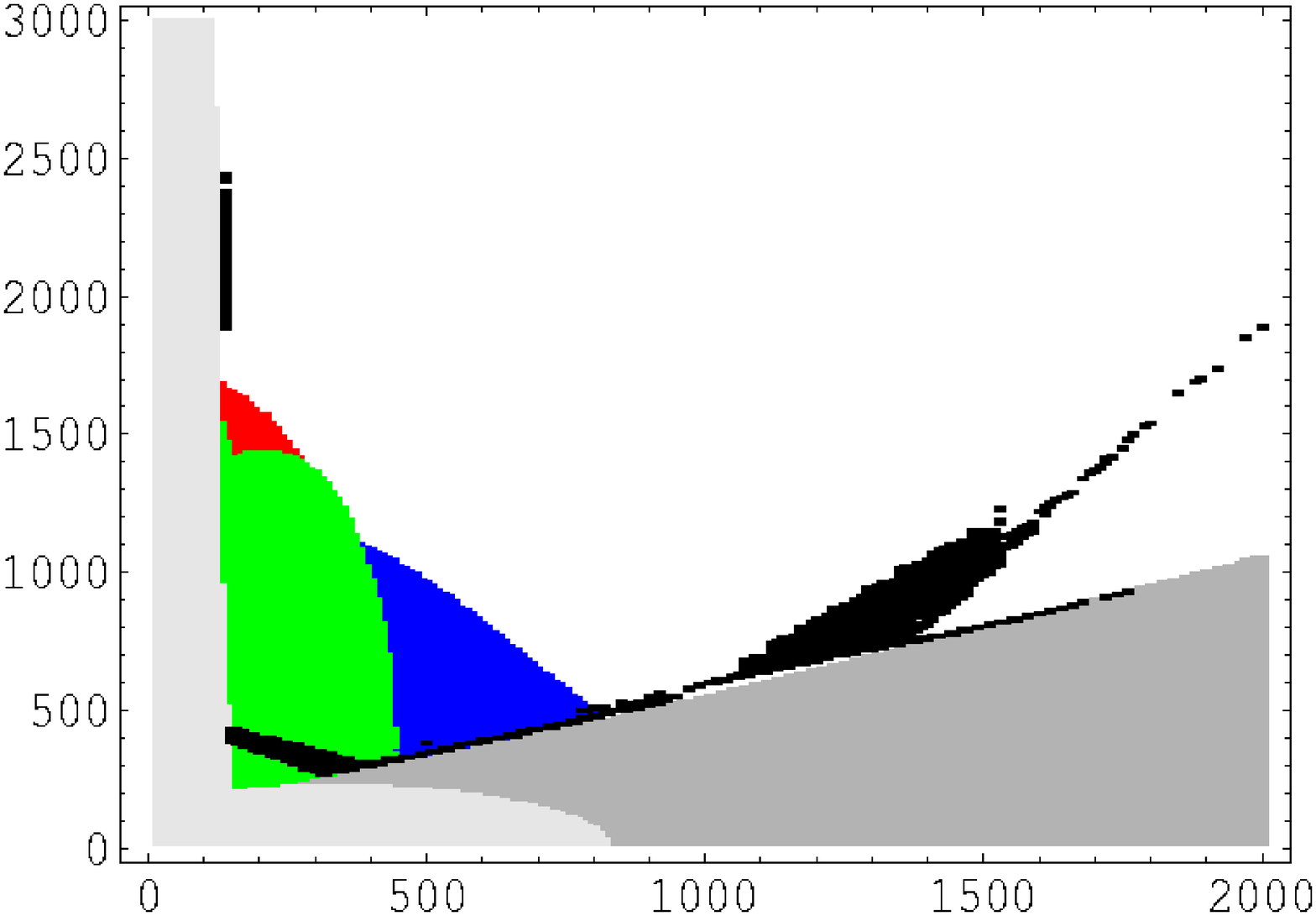,width=7.5cm,height=7.5cm} }
\hspace*{13.9cm} ${\mathbf m_{1/2}}$
\vspace*{-.3cm}
\end{center}
\caption{\it The mSUGRA $(m_{1/2},m_0)$ parameter space with all constraints 
  imposed for $A_0=0, \mu>0$ and the values $\tan\beta=10$ (left) and 50
  (right). The top quark mass is fixed to $m_t=178$ GeV. Notation and
  conventions are as in Fig.~1.}
\vspace*{-.5cm}
\end{figure}

We see that the Higgs search limits are very severe for small and moderate
values of $\tan\beta$, and/or for smaller values of $m_t$; for example, for
$m_t = 172.7$ GeV, $\tan\beta = 5$ and $A_0 = 0$ (Fig.~2) they imply $m_{1/2}
\gsim 0.6$ TeV for small $m_0$, or $m_0 \gsim 2$ TeV for small $m_{1/2}$. At
first the region excluded by this constraint shrinks quickly with increasing
$\tan\beta$, but it becomes almost independent of this parameter once
$\tan\beta \gsim 20$. These regions become even larger if $m_t$ is below its
current central value. For example, Fig.~3 shows that the region excluded by
this constraint for $\tan\beta = 10$ and $m_t = 166.9$ GeV, the current 95\%
c.l.  lower bound on this quantity, is very similar to that for $\tan\beta =
5$ and $m_t$ at its current central value. Conversely, again for $\tan\beta =
10$, increasing $m_t$ to the previous central value of 178 GeV \cite{d0top}
(which happens to be quite close to the current 95\% c.l. upper bound) reduces
the lower bound on $m_{1/2}$ for small $m_0$ to about 250 GeV, and allows
values of $m_0$ down to about 750 GeV even if $m_{1/2}$ is small, see Fig.~4.
All these numbers allow for a 3 GeV theoretical uncertainty in $m_h$.

The (green) regions excluded by the $b \rightarrow s \gamma$ constraints shows
the opposite dependence on $\tan\beta$, becoming larger as this parameter
increases. Note, however, that for $m_t = 172.7$ GeV and $A_0 = 0$ (Figs.~1
and 2) the Higgs constraint supersedes the $b \rightarrow s \gamma$ constraint
for values of $\tan\beta$ up to $\tan\beta = 50$, unless one assumes that
unknown higher--order corrections to $m_h$ are large and positive. For the
highest value displayed, $\tan\beta=58$, the $b \rightarrow s \gamma$
constraint excludes an additional domain close to the region where electroweak
symmetry breaking does not take place; in this area, the charged Higgs boson
is relatively light and the $tH^\pm$ contribution (which is not compensated by
the $\tilde t \tilde{\chi}^\pm$ ones, as the top squarks are rather heavy)
leads to a value of $B(b \rightarrow s \gamma)$ that is slightly higher than
the upper bound $B_{\rm max}=4.45 \cdot 10^{-4}$.

Figs.~3 and 4 show that reducing (increasing) $m_t$ further reduces (increases)
the importance of the $b \rightarrow s \gamma$ constraint relative to the Higgs
search constraint. Taking sizable (negative) values of $A_0$ also increases the
relative importance of the $b \rightarrow s \gamma$ constraint, as shown by
Figs.~5 and 6.\footnote{The effects of $A_0 \neq 0$ have recently been studied
in \cite{pauss}.} 

  For example, for $\tan\beta = 30$ and $A_0 = -2$ TeV (Fig.~6)
the region excluded by the $b \rightarrow s \gamma$ constraint even excludes
some combinations of parameters with predicted $m_h$ well above 115 GeV. The
predicted branching ratio for $b \rightarrow s \gamma$ decays is quite
sensitive to $A_0$ since the main supersymmetric contribution comes from
stop--chargino loops, and stop mixing is affected quite strongly by $A_0$, as
long as $|A_0| \gsim m_{1/2}$. Recall, however, that this constraint can be
more easily circumvented than the other constraints discussed here.

The excluded grey regions also increase with increasing $\tan\beta$. Increased
$\tilde \tau_L - \tilde \tau_R$ mixing as well as reduced diagonal $\tilde \tau$
masses due to RG effects imply that the requirement $m_{\tilde \tau_1} >
m_{\tilde\chi_1^0}$ excludes an increasing area with $m_{1/2}^2 \gg m_0^2$.
Similarly, the requirement of correct electroweak symmetry breaking excludes an
increasing area at $m_0^2 \gg m_{1/2}^2$. For small values of $\tan\beta$ the
increase of this excluded region is mostly due to the reduction of the top
Yukawa coupling, which scales like $1/\sin\beta$; for larger $\tan\beta$ the
effects of the bottom Yukawa, which scales like $1/\cos\beta \simeq \tan\beta$
for $\tan^2\beta \gg 1$, in the RGE become more important, which to some extend
counter--act the contributions of the top Yukawa coupling. For the large value
$\tan\beta=58$ (Fig.~2), this region covers most of the parameter space as the
bottom Yukawa coupling becomes very large. For even larger values, $\tan\beta>
58$, one cannot obtain correct electroweak symmetry breaking.  

The grey regions also depend very sensitively on the top mass, as
shown by Figs.~3 and 4. Figs.~5 and 6 show that increasing $|A_0|$ reduces the
region at $m_0^2 \gg m_{1/2}^2$ where one cannot break the electroweak
symmetry; on the other hand, a significant region near the origin is now
excluded by the false vacuum (``CCB'') constraints.

The (blue) region favored by the measurement of $g_\mu - 2$ (if the SM
prediction using data from $e^+e^-$ annihilation can be trusted) also expands
towards larger values of $m_0$ and $m_{1/2}$ as $\tan\beta$ is increased. This
region depends only slightly on $m_t$; somewhat larger values of $m_0$ become
compatible with this constraint if $m_t$ is reduced. This is due to the
reduction of $\mu$ caused by reducing $m_t$. The dependence of this region on
$A_0$ is again rather mild; however, for $A_0 = -2$ TeV the entire blue region
is excluded by the $b \rightarrow s \gamma$ constraint.

\begin{figure}[h!]
${\mathbf m_0}$
\vspace*{-.3cm}
\begin{center}
\mbox{\epsfig{file=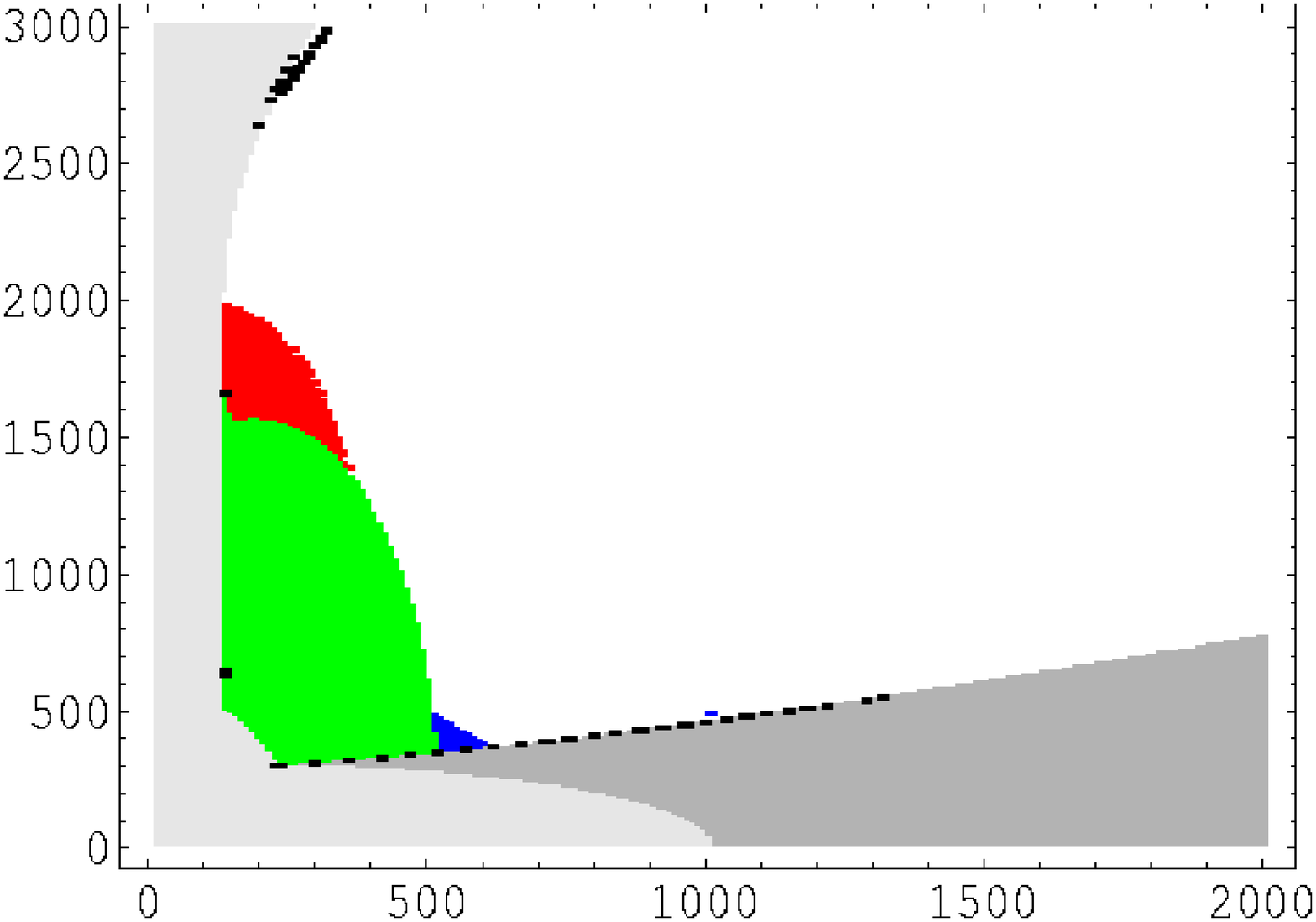,width=7.5cm,height=7.5cm}\hspace*{0.9cm}
      \epsfig{file=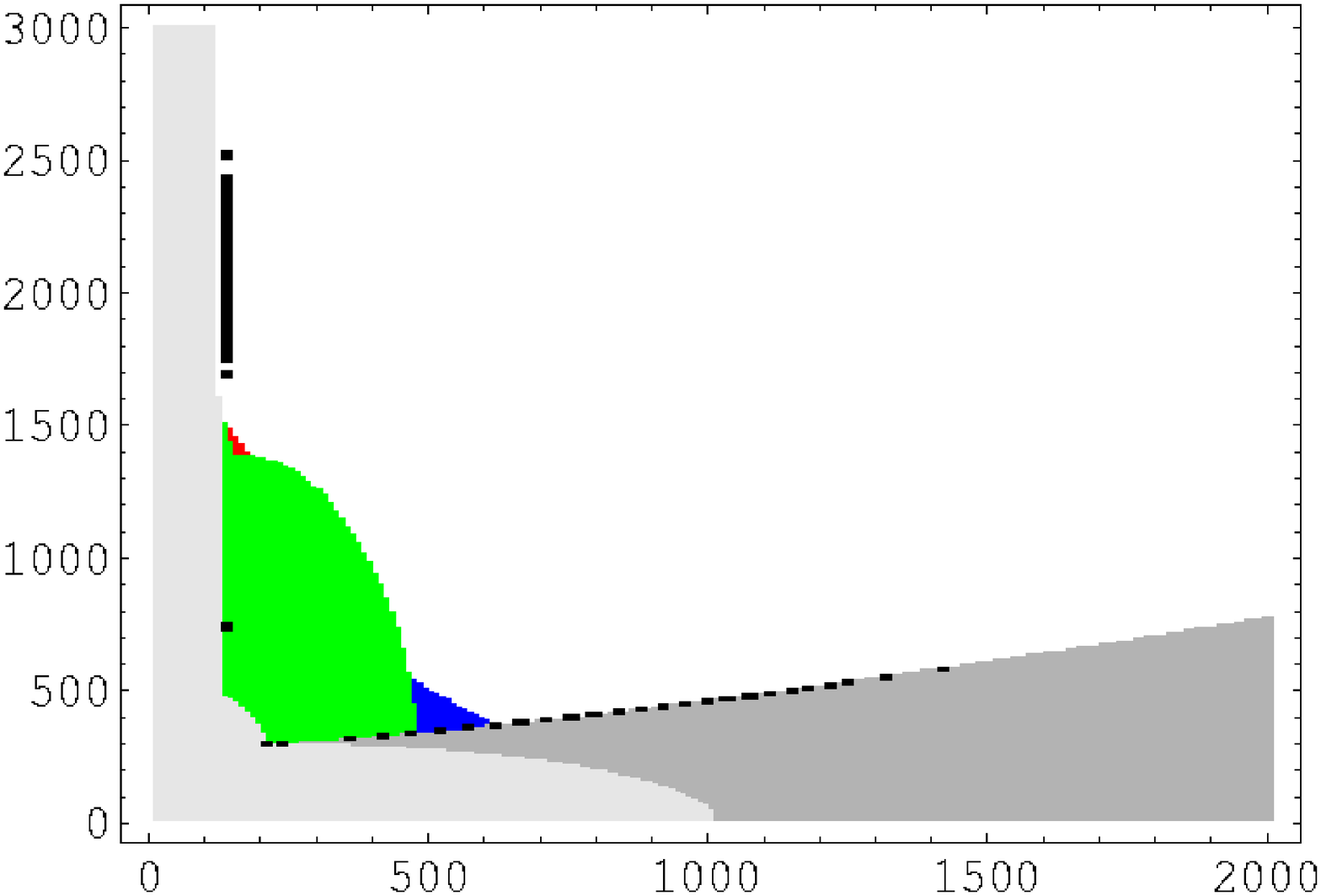,width=7.5cm,height=7.5cm} }
\hspace*{13.9cm} ${\mathbf m_{1/2}}$
\vspace*{-.3cm}
\end{center}
\caption{\it The mSUGRA $(m_{1/2},m_0)$ parameter space with all constraints 
  imposed for $A_0=-1$ TeV, $\mu>0, \tan\beta=30$. The top quark mass is fixed
  to $m_t=172.7$ (left) and 178 GeV (right). Notation and conventions are as in
  Fig.~1.}
\vspace*{.3cm}
\end{figure}

\begin{figure}[h!]
${\mathbf m_0}$
\vspace*{-.3cm}
\begin{center}
\mbox{\epsfig{file=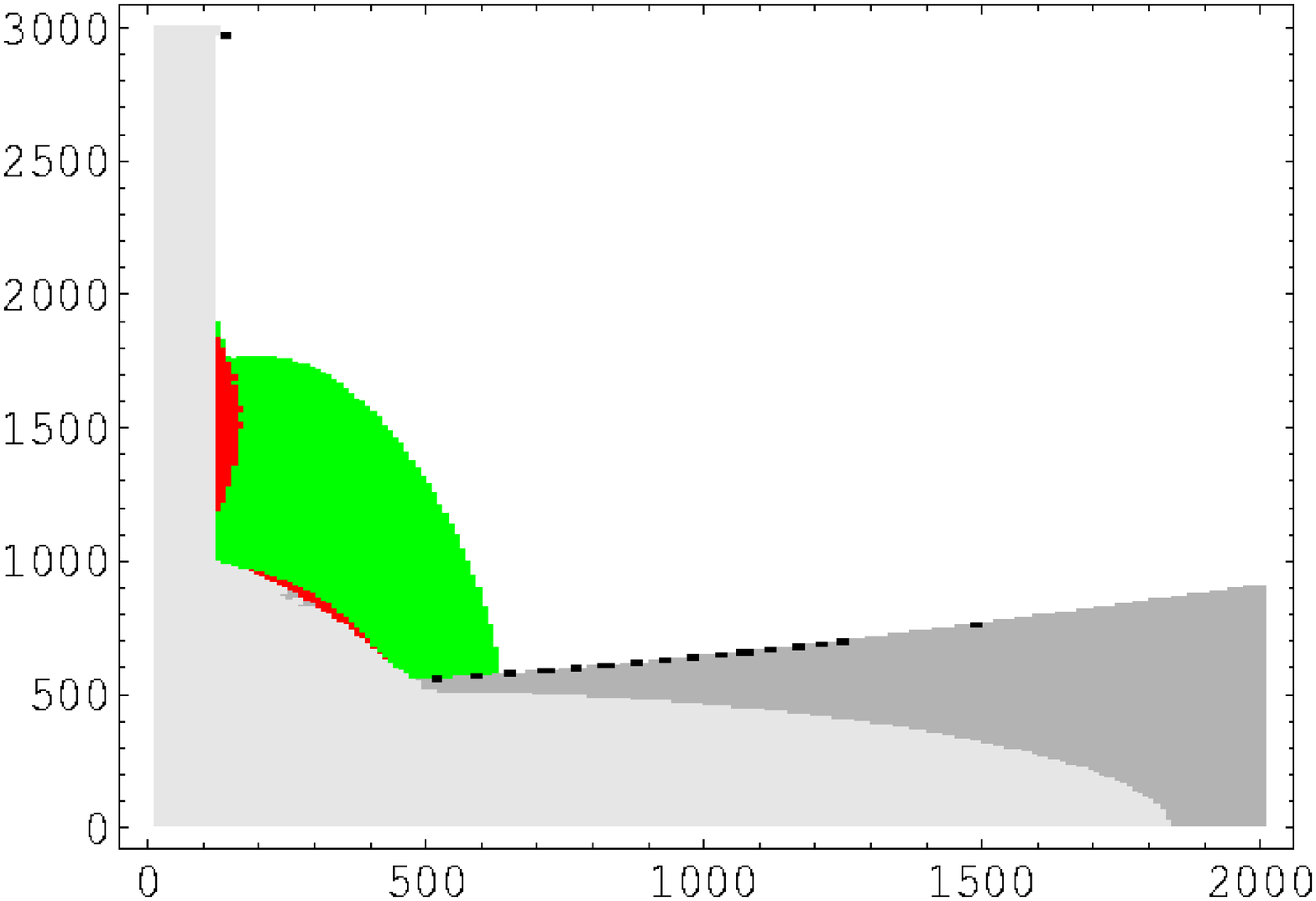,width=7.5cm,height=7.5cm}\hspace*{0.9cm}
      \epsfig{file=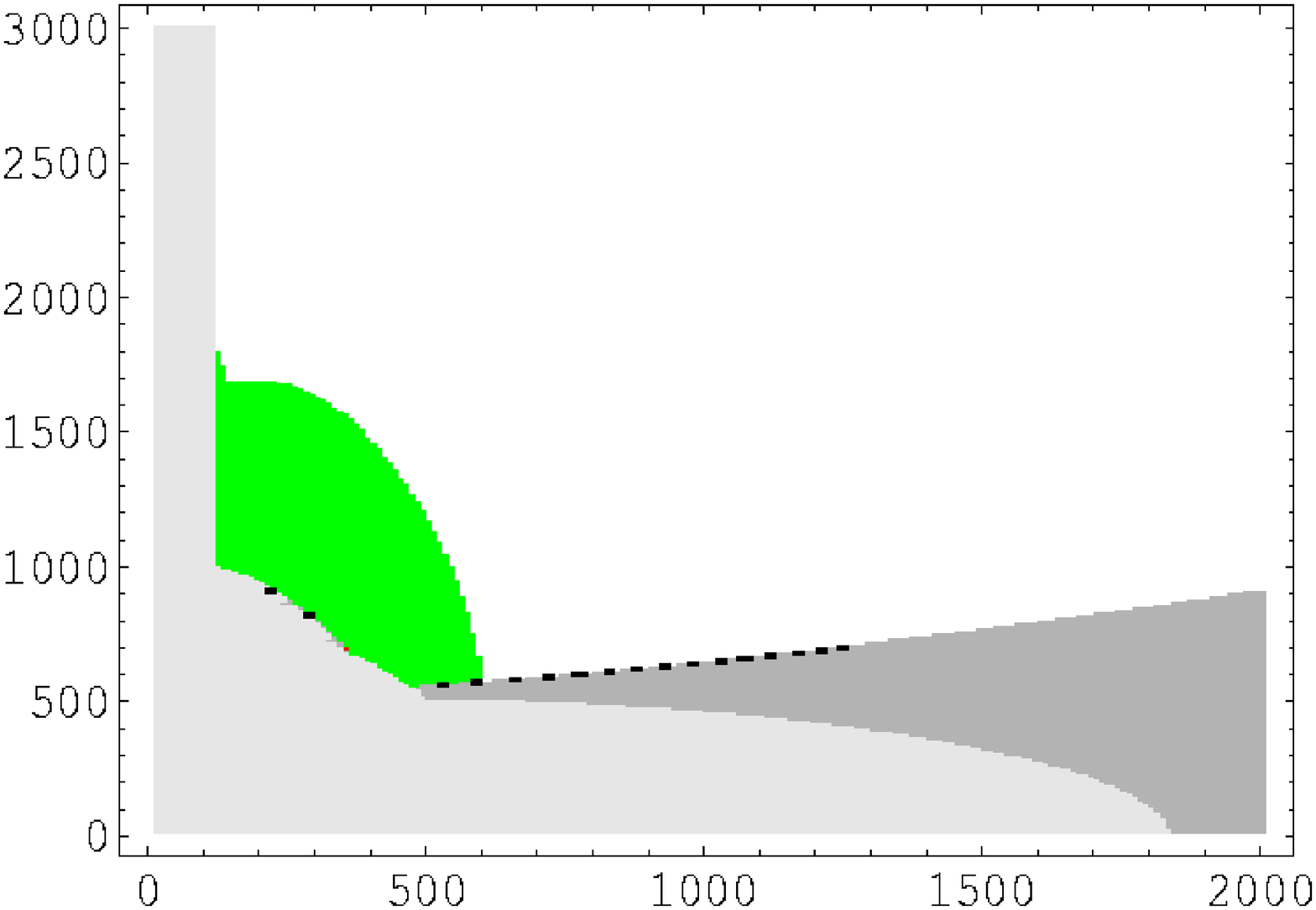,width=7.5cm,height=7.5cm} }
\hspace*{13.9cm} ${\mathbf m_{1/2}}$
\vspace*{-.3cm}
\end{center}
\caption{\it The mSUGRA $(m_{1/2},m_0)$ parameter space with all constraints 
  imposed for $A_0=-2$ TeV, $\mu>0, \tan\beta=30$. The top quark mass is fixed
  to $m_t=172.7$ (left) and 178 GeV (right). Notation and conventions are as in
  Fig.~1.}
\vspace*{-.5cm}
\end{figure}

Finally, for $\tan\beta=5$ (Fig.~2), the (black) regions satisfying the DM
constraint (\ref{dm}) lie right at the border of the theoretically allowed
parameter space: the stau co--annihilation region, where $m_{\tilde \tau_1}
\simeq m_{\tilde \chi_1^0}$, lies next to the region excluded by a charged LSP,
whereas for small $\tan\beta$ the ``focus point region'', where $\mu \lsim M_1$
at the weak scale, is right next to the region where the electroweak symmetry
can no longer be broken. The same holds true for $\tan\beta=10$ and $m_t = 178$
GeV (Fig.~4), for $\tan\beta = 30$ and $A_0 = -2$ TeV (Fig.~6), and for
$\tan\beta = 30, \ A_0 =-1$ TeV and $m_t = 178$ GeV (Fig.~5); in this last case
the small black region at $m_0 \gg m_{1/2}$ is allowed due to almost resonant
$h$ exchange \cite{hpole}.

However, for larger $\tan\beta$ and/or smaller $m_t$ there are sizable regions
of parameter space where the DM relic density comes out too low. This happens
in particular for $m_0^2 \gg m_{1/2}^2$, where one can have an almost pure
higgsino as LSP once $\tan\beta \gsim 10$ (for $m_t = 172.7$ GeV; for smaller
$m_t$ this region appears at smaller $\tan\beta$). Comparison of Figs.~1--2
and 3 shows that the lowest values of $m_0$ in this higgsino--LSP region
decreases very quickly with decreasing $m_t$. The reason is that larger $m_t$
imply a smaller (often: more negative) soft breaking contribution to the
squared mass of the Higgs boson that couples to the top quark, which in turn
implies a larger value of $|\mu|$ via the conditions of electroweak symmetry
breaking.

For the same reason, a reduced $m_t$ makes it easier to find solutions with
$m_A \simeq 2 m_{\tilde \chi_1^0}$, yielding the very prominent DM--allowed
``$A-$pole'' region visible for $\tan\beta = 50$ in Figs.~2--4; recall that
$\mu^2$ contributes with positive sign to $m_A^2$. This region becomes broader
with reduced $m_t$ since lower values of $|\mu|$ also mean stronger
gaugino--higgsino mixing in the neutralino sector, and hence a larger
$\tilde\chi_1^0 \tilde\chi_1^0 A$ coupling. For $\tan\beta=58$ (Fig.~2) $m_A$
is everywhere significantly below $2 m_\chi$. The very large $(A,H) b \bar b$
couplings nevertheless imply that for moderate values of $m_0$ and $m_{1/2}$,
$\tilde \chi_1^0$ annihilation through virtual $A$ and $H$ exchange has the
right strength. This DM--allowed region is connected to the $A-$pole region at
smaller (although still large) values of $\tan\beta$. For $\tan\beta=58$ and
small $m_0$ and $m_{1/2}$, the virtual $A, \, H-$ exchange contributions even
lead to too small a $\tilde\chi_1^0$ relic density; however, this region of
parameter space is excluded by the $b \rightarrow s \gamma$ constraint.

On the other hand, a reduced top mass of 172.7 GeV also implies that the
``bulk'' regions, where the DM constraint (\ref{dm}) is satisfied due to the
exchange of light sleptons in the $t-$ and $u-$channel, now lies deep in the
region excluded by Higgs searches at LEP. Fig.~4 is a reminder that a bulk
region compatible with all constraints (with the possible exception of the
theoretically somewhat shaky $b \rightarrow s \gamma$ constraint) still exists
for $m_t = 178$ GeV and (sufficiently) large $\tan\beta$. Recall that
increasing $\tan\beta$ increases $\tilde \tau_L - \tilde \tau_R$ mixing, which
in turn increases the $S-$wave LSP annihilation cross section through $\tilde
\tau$ exchange \cite{dn3}.

Note finally that the additional possible region where the DM constraint could
be satisfied, i.e. with co--annihilation of the LSP neutralino with top
squarks \cite{stop-co}, is disfavored in the mSUGRA scenario that we are
discussing here.

It is interesting to note that several indications for ``new physics'' can be
explained simultaneously within mSUGRA. The reduction of the central value of
$m_t$ has made it a bit more difficult to satisfy the (aggressive) $g_\mu - 2$
requirement (\ref{gmuc2}), which prefers moderate values of sparticle masses
unless $\tan\beta$ is quite large, in potential conflict with the LEP Higgs
search limits. However, if we allow a 3 GeV theoretical uncertainty in the
calculation of $m_h$, solutions satisfying (\ref{gmuc2}) can be found for all
$\tan\beta \geq 8$ for $m_t = 172.7$ GeV and $A_0 = 0$; if finite values of
$A_0$ are considered, the lower limit on $\tan\beta$ is reduced even further.
On the other hand, if we take the prediction of $m_h$ at face value, again
taking $m_t = 172.7$ GeV we need $\tan\beta \geq 12$ (7) for vanishing
(arbitrary) $A_0$; for $m_t = 166.9$ GeV, as in Fig.~3, these lower bounds
increase to 20 and 10, respectively. In all these cases we can satisfy the DM
constraint (\ref{dm}) in the $\tilde \tau_1$ co--annihilation region, and have
a CP--even Higgs boson near 115 GeV, as hinted at by LEP data, while
satisfying the $g_\mu-2$ constraint (\ref{gmuc2}) at the same time. 

Fig.~2 shows that these three constraints can also be satisfied simultaneously
in the $A-$pole region, if $\tan\beta$ is very large. However, we did not find
any points in the ``focus point'' region where the aggressive $g_\mu-2$
constraint can be satisfied, if we take the prediction for $m_h$ at face
value. In this case increasing $m_t$, and/or introducing non--vanishing $A_0$,
would allow to satisfy the Higgs search limits for smaller values of $m_0$,
thereby increasing $a_{\mu,\,{\rm SUSY}}$; however, at the same time it would
increase $|\mu|$, pushing the $\tilde\chi_1^0$ relic density to unacceptably
large values. Even if we only demand the calculated $m_h$ to exceed 111 GeV
(i.e. assume sizable positive higher order corrections to $m_h$) one can only
reach the lower end of the range (\ref{gmuc2}) in this region of parameter
space.

\subsection*{3.2 Parameter space with physical masses}

We now present some results for physical masses. In order to keep the number
of figures manageable, we only show results for the central value of $m_t =
172.7$ GeV, $A_0 = 0$, and two values of $\tan\beta$. 

\begin{figure}[h!]
\begin{center}
\mbox{\epsfig{file=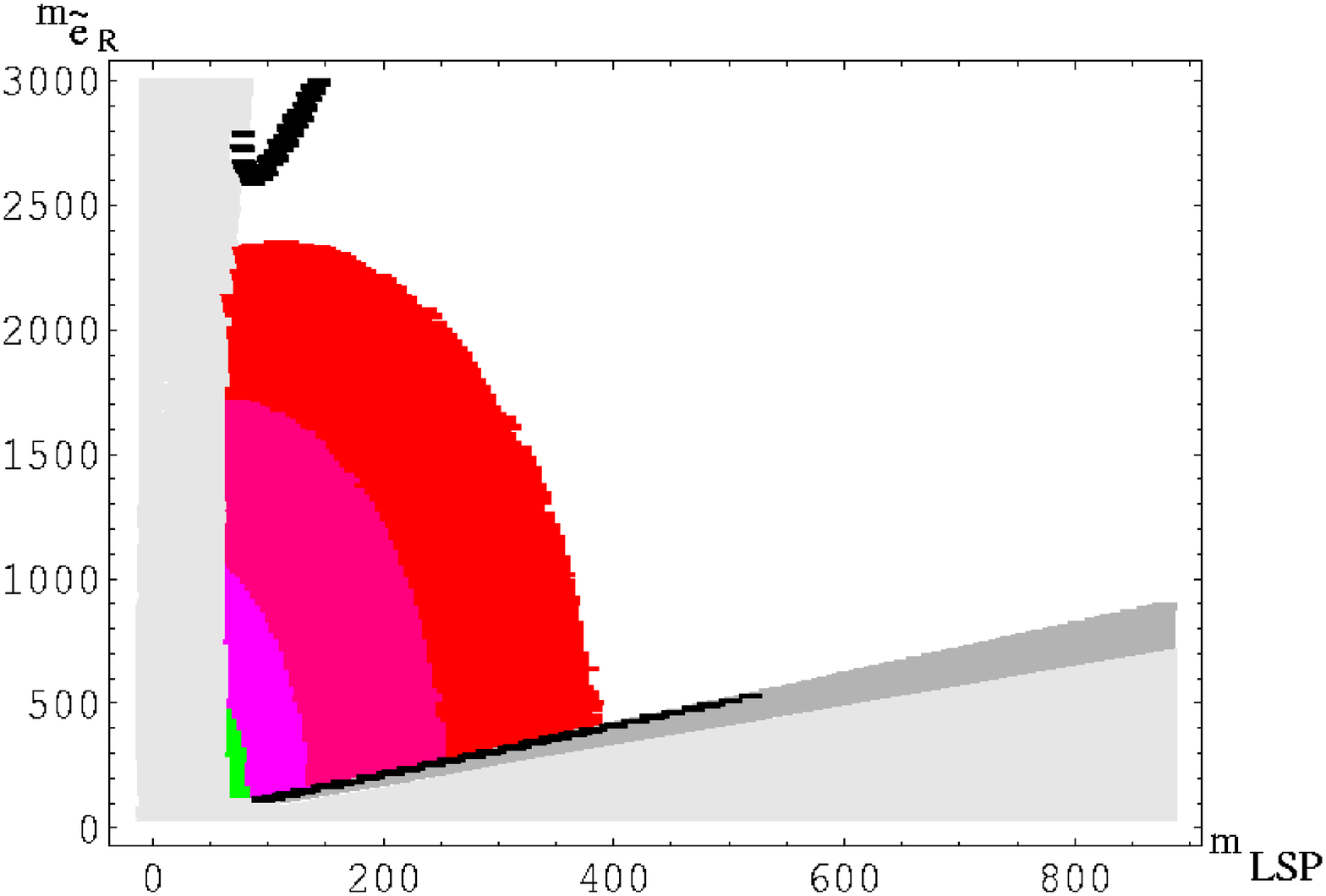,width=7.5cm,height=7.5cm}\hspace*{0.9cm}
      \epsfig{file=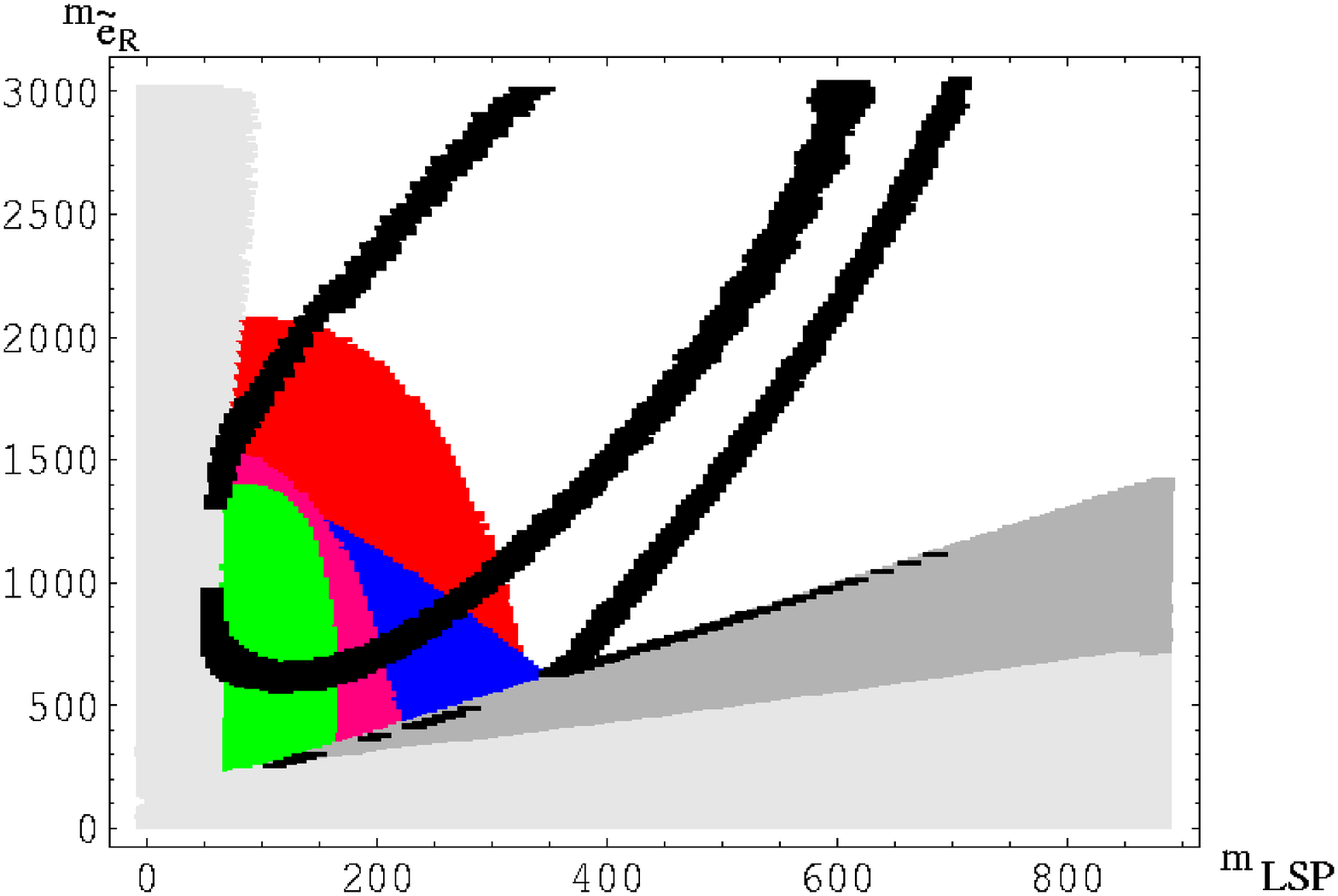,width=7.5cm,height=7.5cm} }
\vspace*{.1cm}
\end{center}
\caption{\it The mSUGRA  $(m_{ {\tilde \chi}_1^0}, m_{ {\tilde e}_R}$) 
  parameter space with all constraints imposed for $A_0=0$, $\mu>0,
  \tan\beta=10$ (left) and $\tan\beta=50$ (right). The top mass is fixed to
  $m_t=172.7$ GeV. Notations and conventions are as in Fig.~1, except that the
  light grey region now also includes combinations of masses that are never
  realized in mSUGRA.}
\end{figure}

We begin with the $(m_{\tilde \chi_1^0}, m_{\tilde e_R})$ plane depicted in
Fig.~7; these masses largely determine the phenomenology of $\tilde e_R$ pair
production at $e^+e^-$ colliders. These figures actually look quite similar to
the corresponding results in Figs.~1--2. The reason is that over most of mSUGRA
parameter space, $\tilde \chi_1^0$ is dominated by the Bino component, with
$m_{\tilde \chi_1^0} \sim 0.4 m_{1/2}$, whereas $m^2_{\tilde e_R} \sim m_0^2 +
0.15 m_{1/2}^2$ implies that the mass of $\tilde e_R$ is usually quite close
to $m_0$. Note, however, that the excluded region at $m^2_{\tilde e_R} \gg
m^2_{\tilde \chi_1^0}$ does not grow with increasing $\tan\beta$, in contrast
to the analogous region in the $(m_0, m_{1/2})$ plane. The reason is that the
boundary\footnote{Most of this region is excluded since electroweak symmetry
  breaking would require $\mu^2 < 0$; however, there is also a small area
  where $\mu^2$ is positive, but below the LEP limit.} of this region is set
by the search limit for (higgsino--like) charginos at LEP, which essentially
fixes the mass of $\tilde \chi_1^0$, which is also higgsino--like here.

The $(m_{\tilde g}, m_{\tilde u_L})$ plane is shown in Fig.~8; since the other
first and second generation squarks have masses quite close to $m_{\tilde
  u_L}$ this plane essentially determines the cross section for the production
of strongly interacting sparticles at hadron colliders \cite{book} (with the
possible exception of a light $\tilde t_1$; see below). In this case both
masses depend significantly on the gaugino mass parameter, with $m_{\tilde
  u_L}^2 \sim m_0^2 + 6 m_{1/2}^2$ and $m_{\tilde g} \sim 2.5 m_{1/2}$. As a
result, the accessible part of parameter space gets squeezed, whereas the
entire region $m_{\tilde u_L} \lsim 0.8 m_{\tilde g}$ is not accessible
\cite{ell}, greatly increasing the size of the grey regions compared to the
analogous results of Figs.~1--2. Moreover, since $m_{\tilde g}$ is independent
of $\mu$, the region at $m^2_{\tilde u_L} \gg m_{\tilde g}^2$ that is excluded
because $\mu^2$ comes out too small does grow with increasing $\tan\beta$.
Note that our basic parameter scan only included values of $m_0$ up to 3 TeV.
As a result, the area with $m_{\tilde u_L} > 3$ TeV and much smaller
$m_{\tilde g}$ did not get probed, although some of it is theoretically
accessible.

\begin{figure}[ht]
\begin{center}
\mbox{\epsfig{file=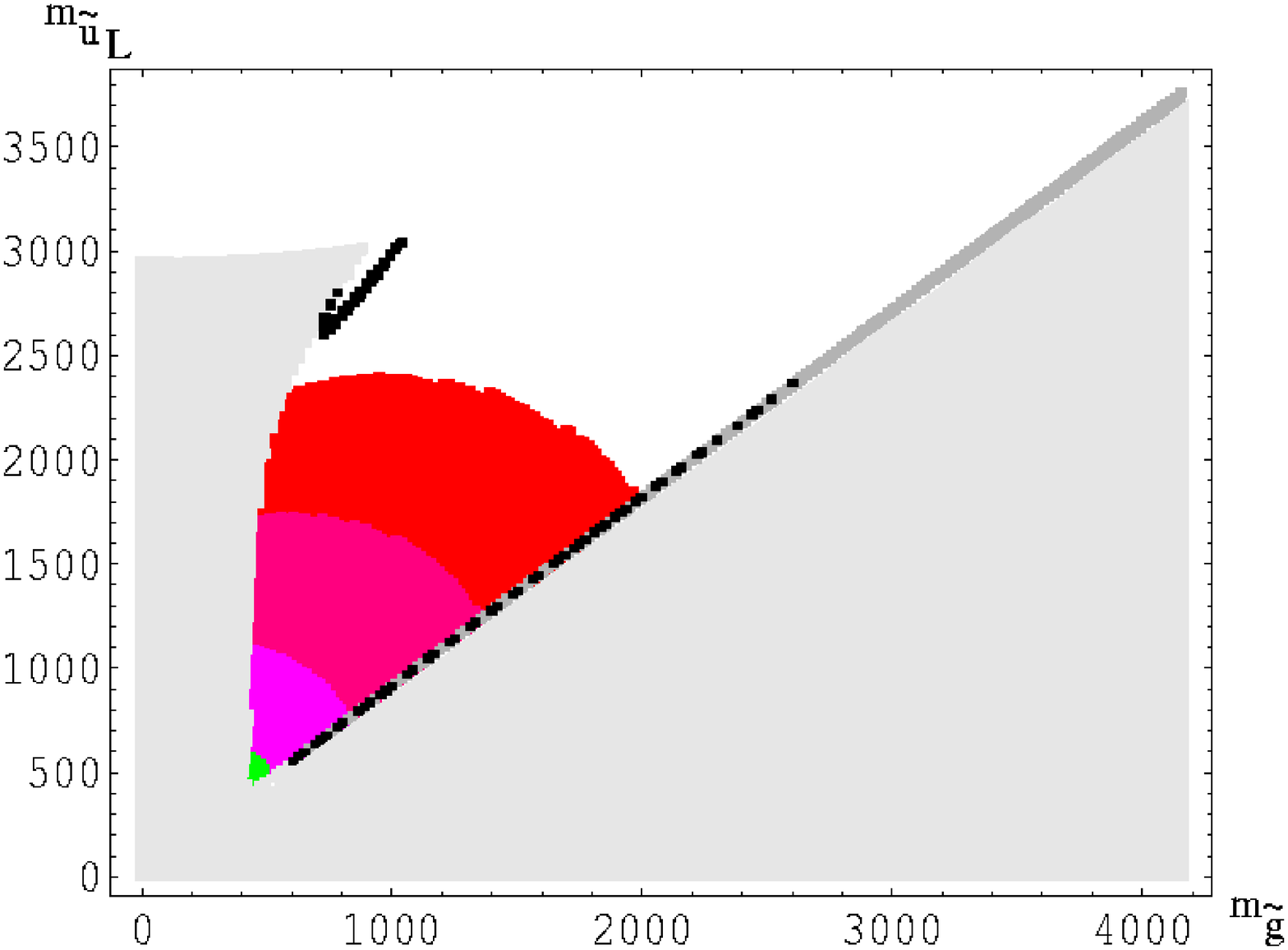,width=7.5cm,height=7.5cm}\hspace*{0.9cm}
      \epsfig{file=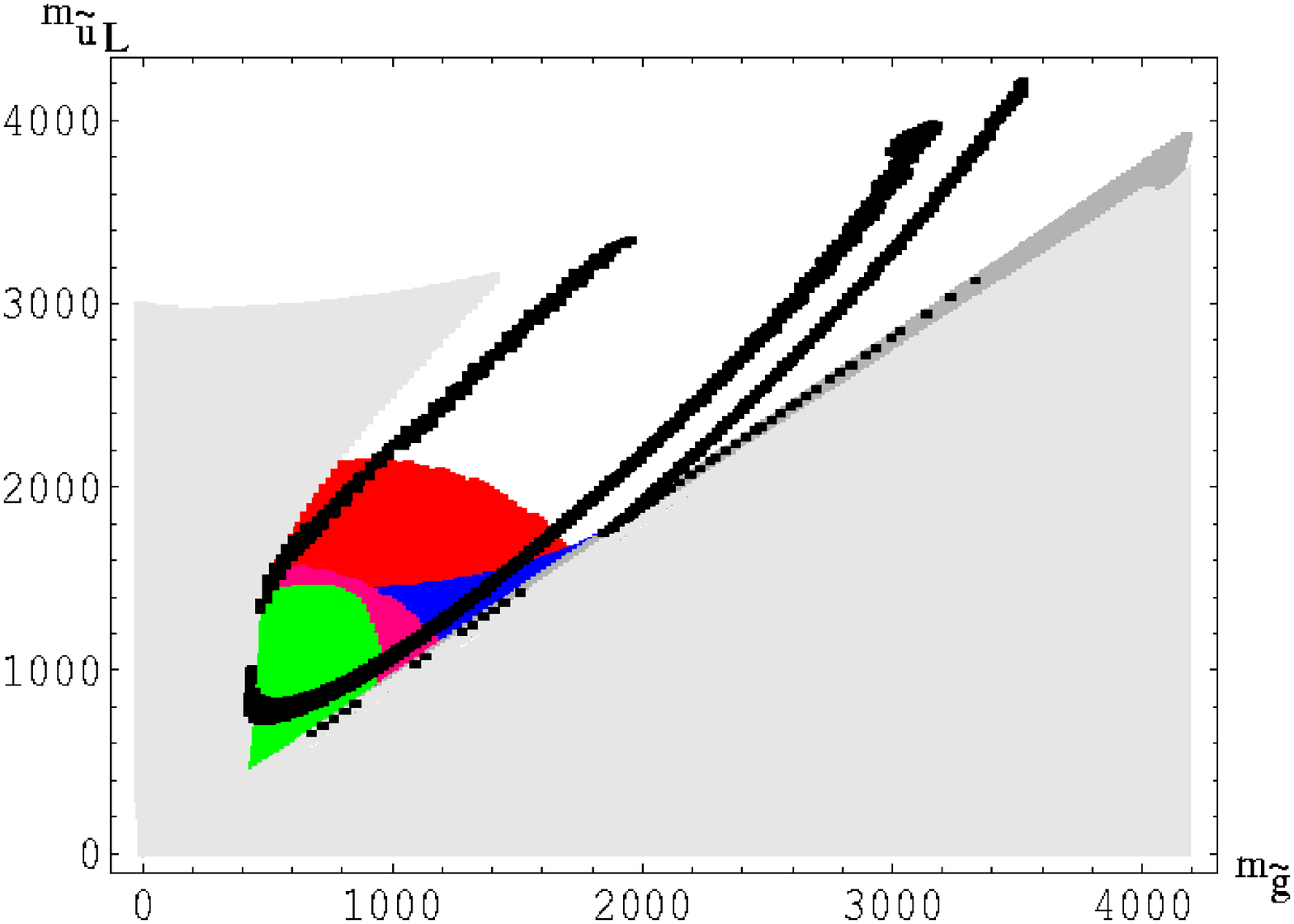,width=7.5cm,height=7.5cm} }
\vspace*{.1cm}
\end{center}
\caption{\it The mSUGRA  $(m_{\tilde g}, m_{\tilde u_L}$) parameter space with 
all constraints imposed for $A_0=0$, $\mu>0, \tan\beta=10$ (left) and 
$\tan\beta=50$ (right). The top mass is fixed to  $m_t=172.7$ GeV.}
\end{figure}

We next turn to the $(m_{\tilde t_1}, m_{\tilde \tau_1})$ parameter space
depicted in Fig.~9. In this case both masses depend significantly on $m_0$,
but only $m_{\tilde t_1}$ has a strong dependence on $m_{1/2}$. The ``focus
point'' region with higgsino--like or mixed LSP is therefore still at the
top--left of the accessible region. However, the inaccessible region at the
left side of the figure is considerably larger than in Figs.~1--2, since for our
choice $A_0 = 0$, one cannot have $m_{\tilde \tau_1}$ too much above
$m_{\tilde t_1}$. Note also that we chose different $y-$scales in the two
frames of Fig.~9.  Increasing $\tan\beta$ increases the $\tau$ Yukawa
coupling, which reduces the soft breaking masses in the $\tilde \tau$ sector
through RG effects, and increases $\tilde \tau_L- \tilde \tau_R$ mixing; both
effects tend to reduce $m_{\tilde \tau_1}$. For $\tan\beta = 10$, $m_{\tilde
  \tau_1}$ is quite close to $m_{\tilde e_R}$, but for $\tan\beta=50$ it is
significantly smaller. On the other hand, $m_{\tilde t_1}$ is relatively
insensitive to $\tan\beta$.

\begin{figure}[ht]
\begin{center}
\mbox{\epsfig{file=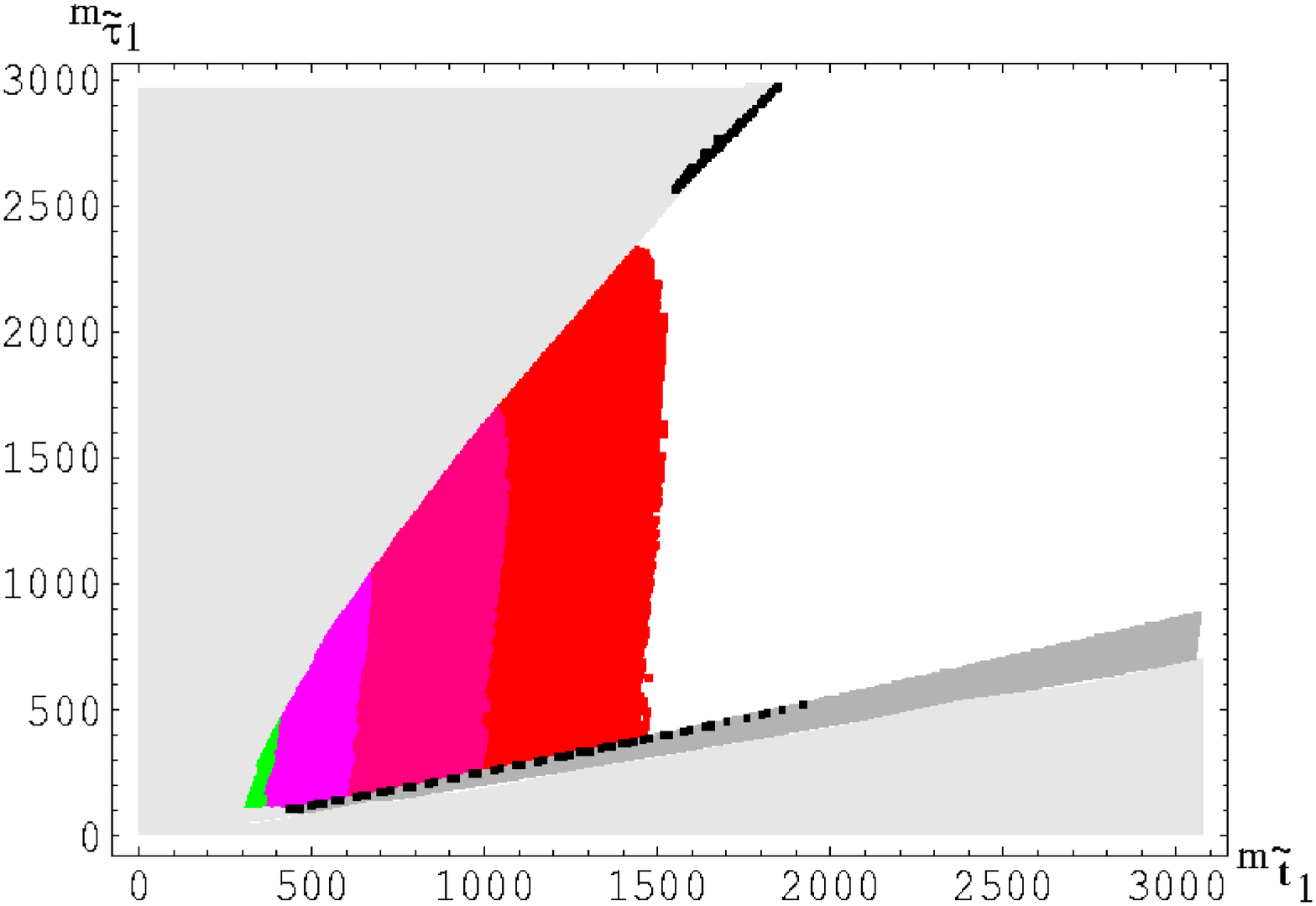,width=7.5cm,height=7.5cm}\hspace*{0.9cm}
      \epsfig{file=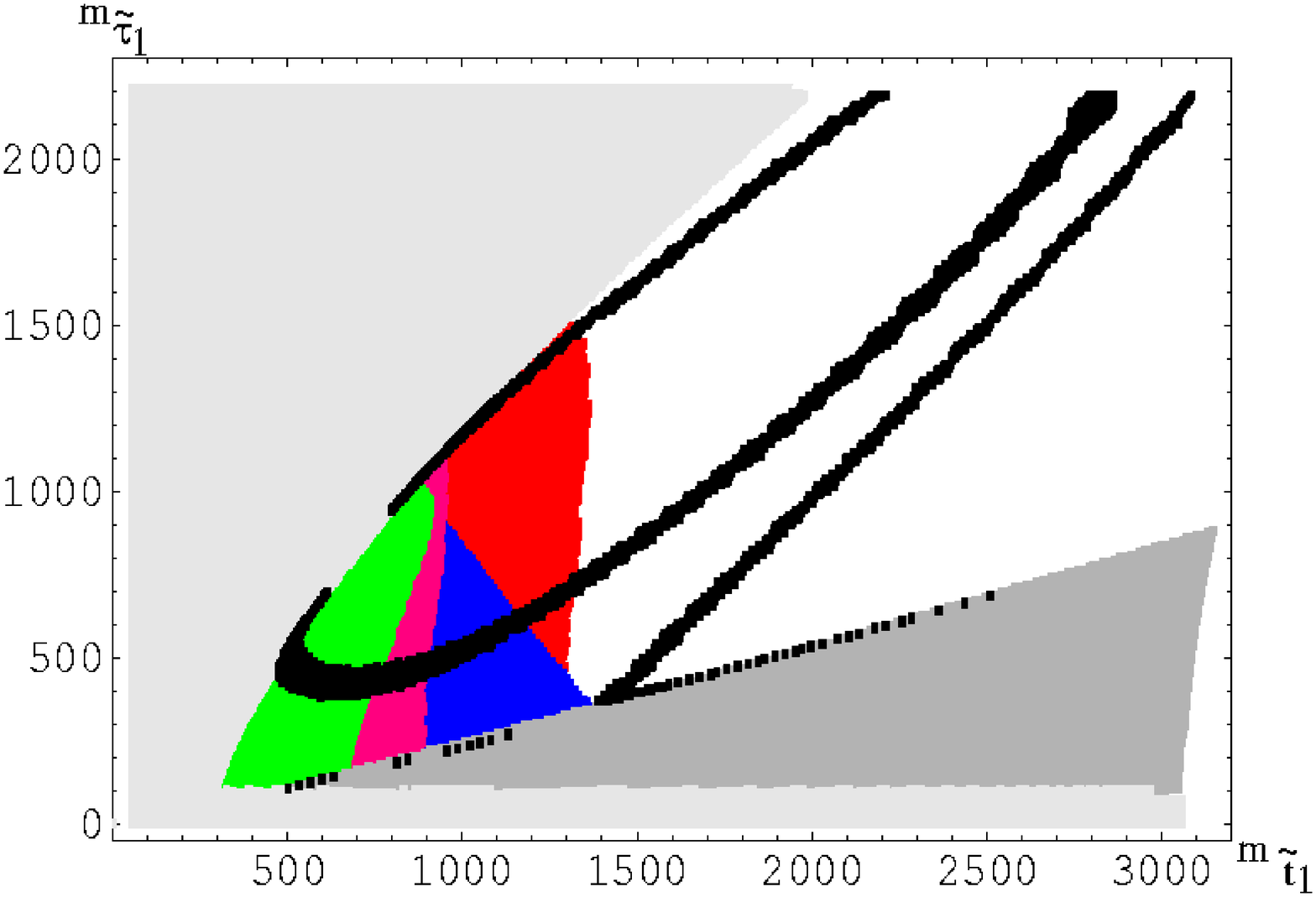,width=7.5cm,height=7.5cm} }
\vspace*{.1cm}
\end{center}
\caption{\it The mSUGRA  $(m_{ {\tilde t}_1}, m_{ \tilde{\tau}_1}$)
  parameter space with all constraints imposed for $A_0=0$, $\mu>0,
  \tan\beta=10$ (left) and $\tan\beta=50$ (right). The top mass is fixed to
  $m_t=172.7$ GeV.} 
\end{figure}
\begin{figure}[h!]
\begin{center}
\mbox{\epsfig{file=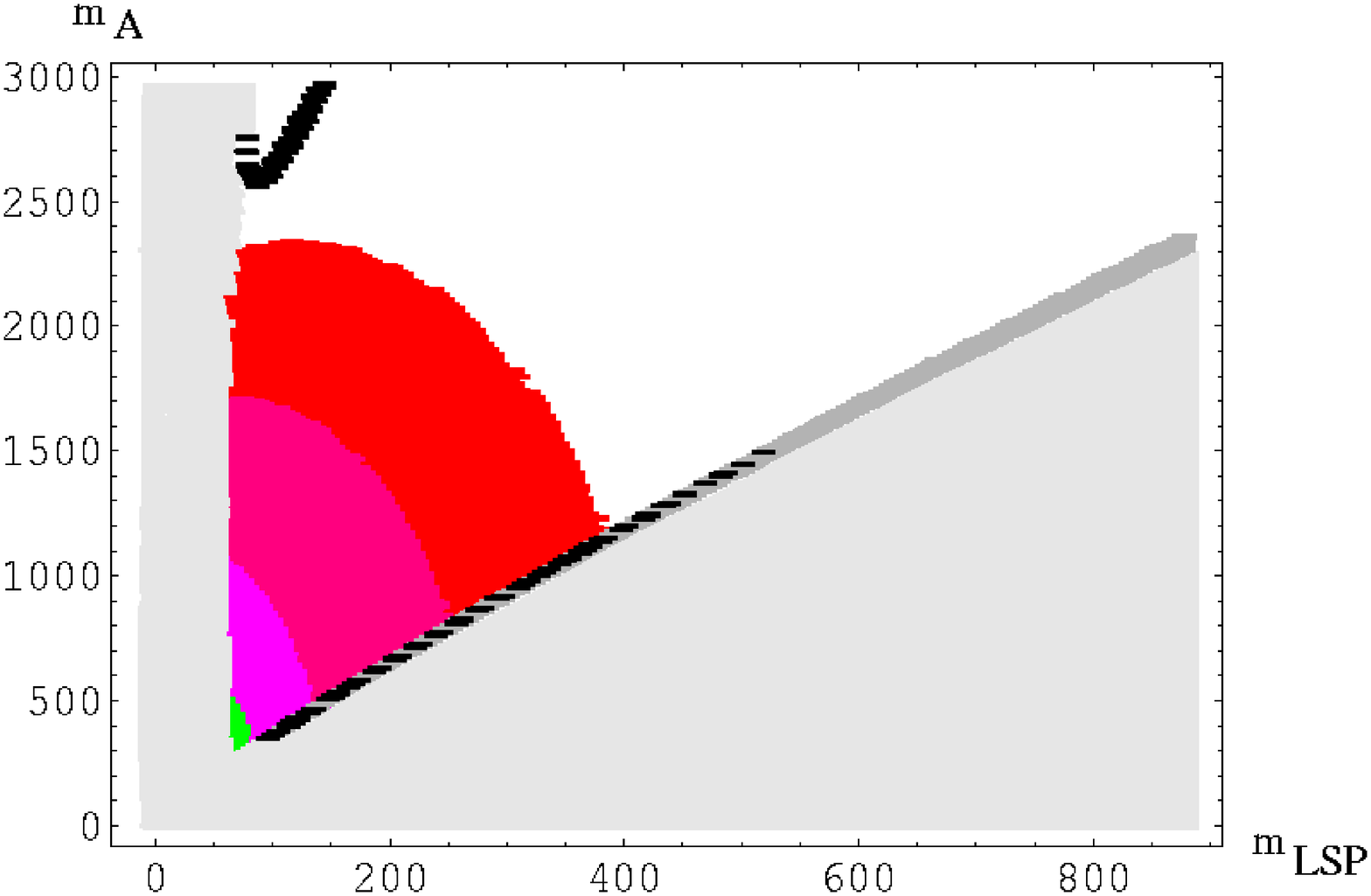,width=7.5cm,height=7.5cm}\hspace*{0.9cm}
      \epsfig{file=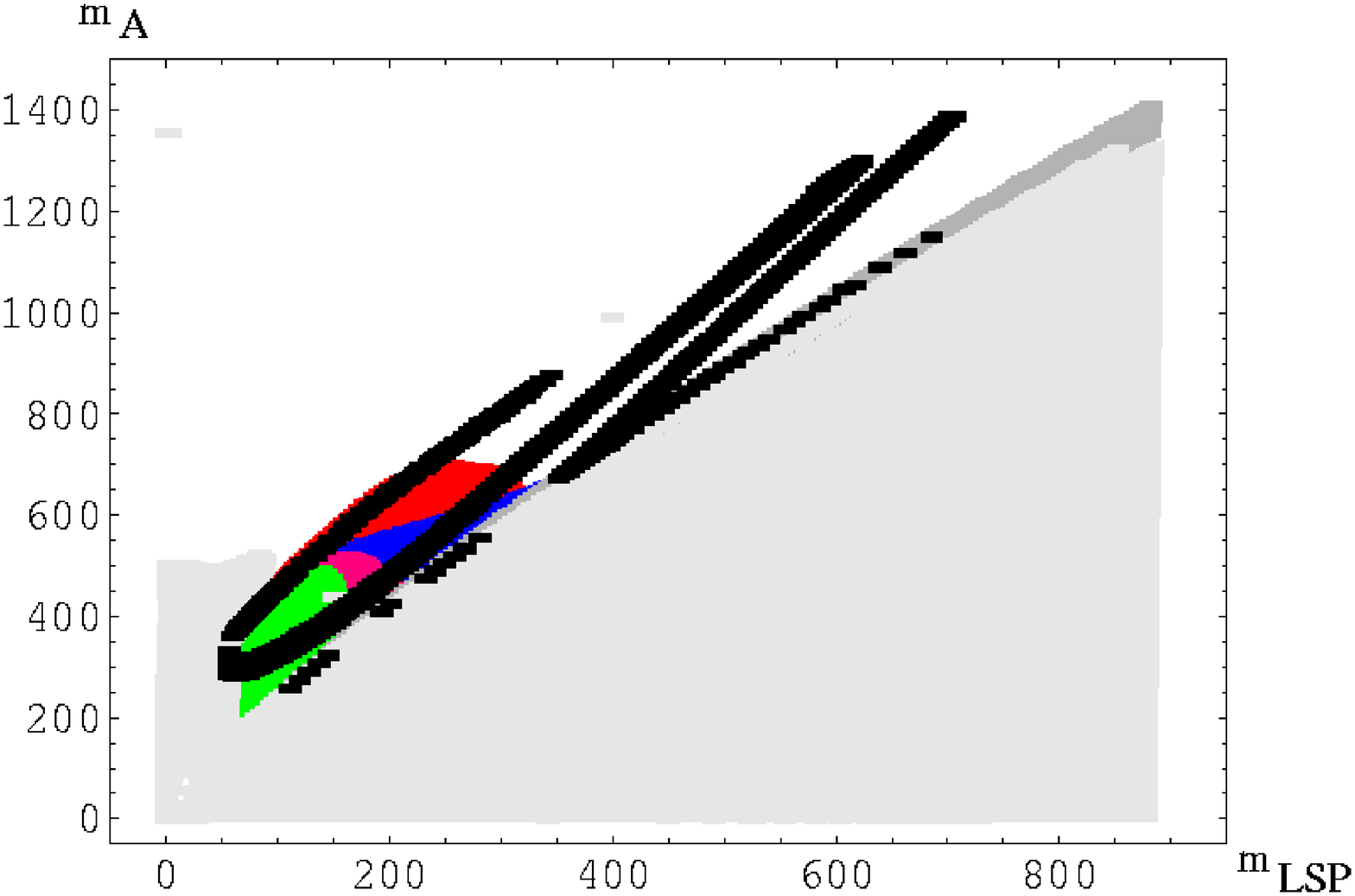,width=7.5cm,height=7.5cm} }
\vspace*{.1cm}
\end{center}
\caption{\it The mSUGRA  $(m_{ {\tilde \chi}_1^0}, m_A$) parameter space
  with all constraints imposed for $A_0=0$, $\mu>0, \tan\beta=10$ (left) and 
$\tan\beta=50$ (right). The top mass is fixed to  $m_t=172.7$ GeV.}
\end{figure}

The Higgs sector \cite{abrev} reflects the radiative symmetry breaking in
mSUGRA. For small and moderate values of $\tan\beta$ the heavier Higgs bosons,
whose masses are essentially determined by that of the CP--odd Higgs boson
$A$, are among the heaviest of all new particles \cite{dn2}. On the other
hand, for large $\tan\beta$, RG effects due to the bottom Yukawa coupling
greatly reduce $m_A$.  We saw in the previous section that this leads to
scenarios where $m_A \simeq 2 m_{\tilde \chi_1^0}$, and hence to strong
$\tilde \chi_1^0$ annihilation in the early Universe due to near--resonant $A$
boson exchange. This reduction of $m_A$ is reflected by the reduced $y-$scale
in the right frames in Figs.~10 and 12, which show the $(m_{\tilde \chi_1^0},
m_A)$ and $(m_h,m_A)$ planes, respectively. In fact, Fig.~10 shows that for
$\tan\beta = 10$, $m_A$ is at least 2.5 times larger than the LSP mass, and
could be arbitrarily large if very large values of $m_0$ are included in the
scan. On the other hand, for $\tan\beta = 50$, only a very narrow range of
$m_A$ values is possible for a given LSP mass, indicating that now $m_A$ has
become almost (although not quite) independent of $m_0$.

The $(m_{\tilde \chi_1^0}, m_h)$ plane is shown in Fig.~11. In the left frame
we see that the upper black (DM--allowed) region is also almost horizontal,
indicating that in this ``focus point'' region $m_h$ depends very weakly on
$m_{1/2}$. In fact, since we have $m_A^2 \gg m_h^2$ everywhere in this frame,
$m_h$ depends on $m_0$ and $m_{1/2}$ only through loop corrections, in
particular through the soft breaking stop masses, which are not sensitive to
$m_{1/2}$ as long as $m_0^2 \gg m_{1/2}^2$. Since in Fig.~1 this region covers
only a narrow range of $m_0$ it also corresponds to a narrow range of $m_h$.
On the other hand, in the $\tilde \tau_1$ co--annihilation region we have
$m_0^2 \ll m^2_{1/2}$, so that the $\tilde t$ masses are mostly determined by
$m_{1/2}$. Since this region covers a broad range of $m_{1/2}$, it also
extends over a significant range of $m_h$. In this second branch the
logarithmic increase of $m_h$ with the SUSY mass scale, here represented by
the LSP mass, is clearly visible.

\begin{figure}[h!]
\begin{center}
\mbox{\epsfig{file=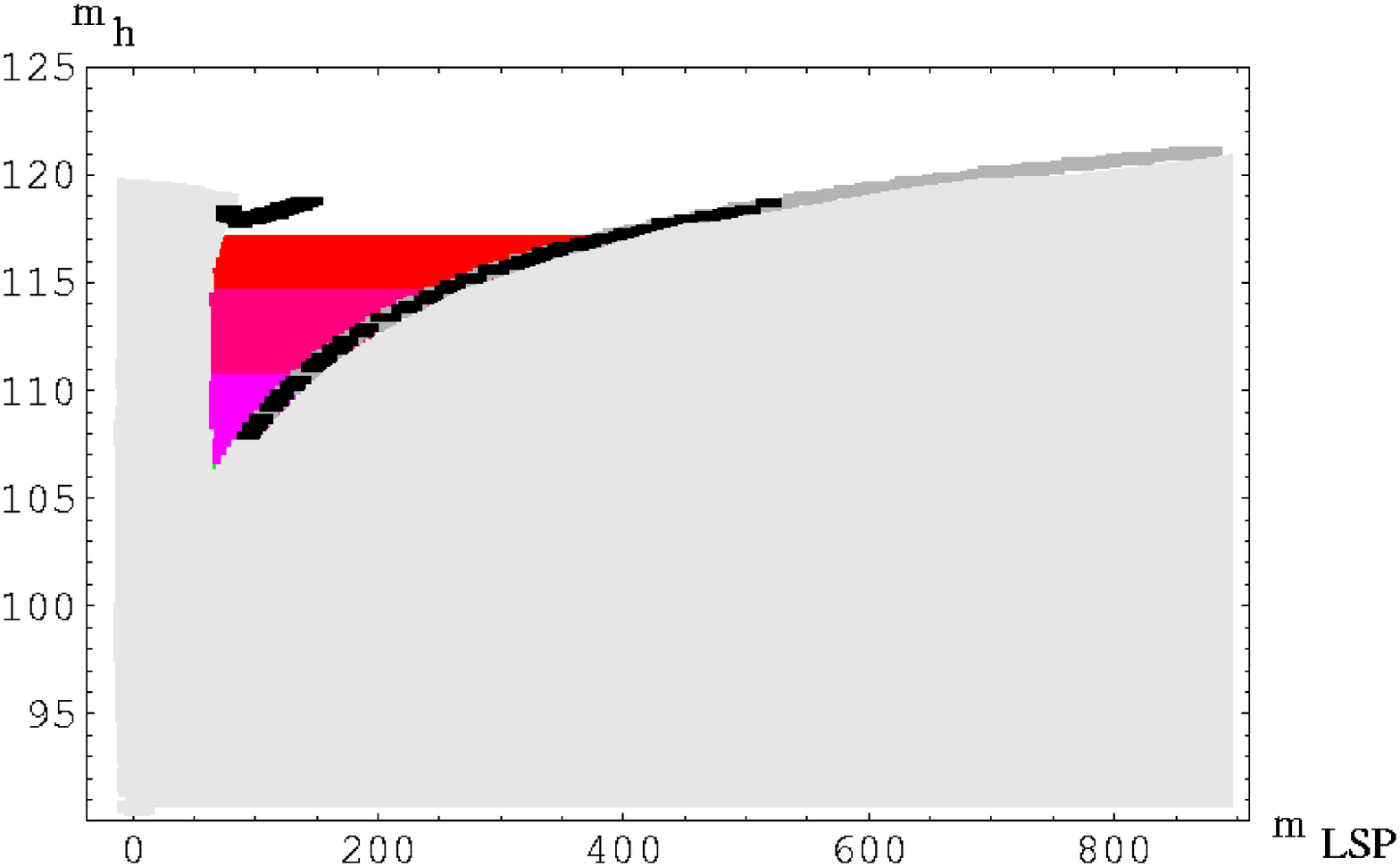,width=7.5cm,height=7.5cm}\hspace*{0.9cm}
      \epsfig{file=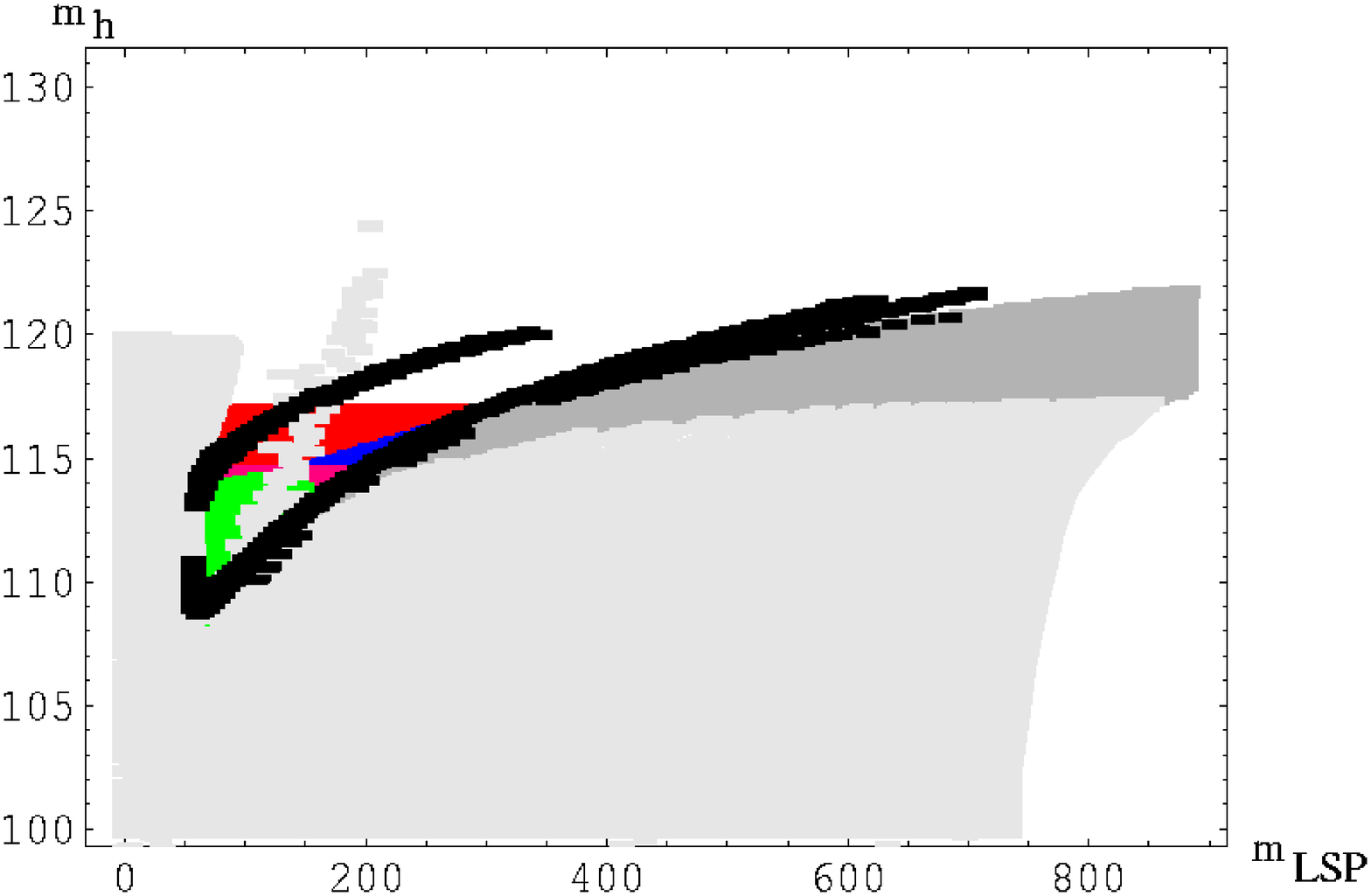,width=7.5cm,height=7.5cm} }
\vspace*{.1cm}
\end{center}
\caption{\it The mSUGRA  $(m_{ {\tilde \chi}_1^0}, m_h$) parameter space
  with all constraints imposed for $A_0=0$, $\mu>0, \tan\beta=10$ (left) and 
$\tan\beta=50$ (right). The top mass is fixed to  $m_t=172.7$ GeV.}
\end{figure}

For $\tan\beta=50$ the four distinct DM--allowed bands shown in the
corresponding Figs.~2 or 7 collapse to two bands. The $\tilde \tau_1$
co--annihilation band is connected to the band where $2 m_{\tilde \chi_1^0}$
is slightly above $m_A$ even in the $(m_0, m_{1/2})$ plane. Together with the
band where $2 m_{\tilde \chi_1^0}$ is slightly below $m_A$, which gives very
similar results for $m_h$, they form the lower black band in Fig.~11. The
DM--allowed region with mixed higgsino--bino LSP, which is quite prominent for
$\tan\beta = 50$, gives the upper black band in the right frame of Fig.~11.
Note also that for fixed $m_{\tilde \chi_1^0}$, in the DM--allowed region
$m_h$ for $\tan\beta=50$ exceeds that for $\tan\beta = 10$ only by about 1
GeV. However, in the latter case $m_{\tilde \chi_1^0}$ up to $\sim 700$ GeV
can be compatible with the DM constraint, whereas for $\tan\beta = 10$ this
requires $m_{\tilde \chi_1^0} \lsim 500$ GeV; the upper bound on $m_h$ in the
DM--allowed region therefore grows by about three GeV when $\tan\beta$ is
increased from 10 to 50. These results are compatible with those of
ref.~\cite{nano-higgs}.

Finally, the $(m_h, m_A)$ plane is shown in Fig.~12. The most obvious feature
here is the strong correlation of these two Higgs boson masses. In the left
frame ($\tan\beta = 10$) the lower black (DM--allowed) strip is the $\tilde
\tau_1$ co--annihilation region, whereas the upper black strip is the ``focus
point'' region. In the latter part of parameter space $|\mu|$ is relatively
small, and effects of the bottom Yukawa coupling are still almost negligible,
so that $m_A \simeq m_0$. Moreover, both $m_A$ and $m_h$ are quite insensitive
to $m_{1/2}$, so long as $m_0^2 \gg m_{1/2}^2$. This leads to a strong
compression of the accessible region, which in this part of parameter space
almost coincides with the DM--allowed region. However, the upper black strip
in the right frame is surrounded by ``inaccessible'' light grey regions only
because we limited our scan to $m_0 \leq 3$ TeV; otherwise it would be
connected by the (experimentally and theoretically allowed, but
DM--disfavored) white region.

\begin{figure}[h!]
\begin{center}
\mbox{\epsfig{file=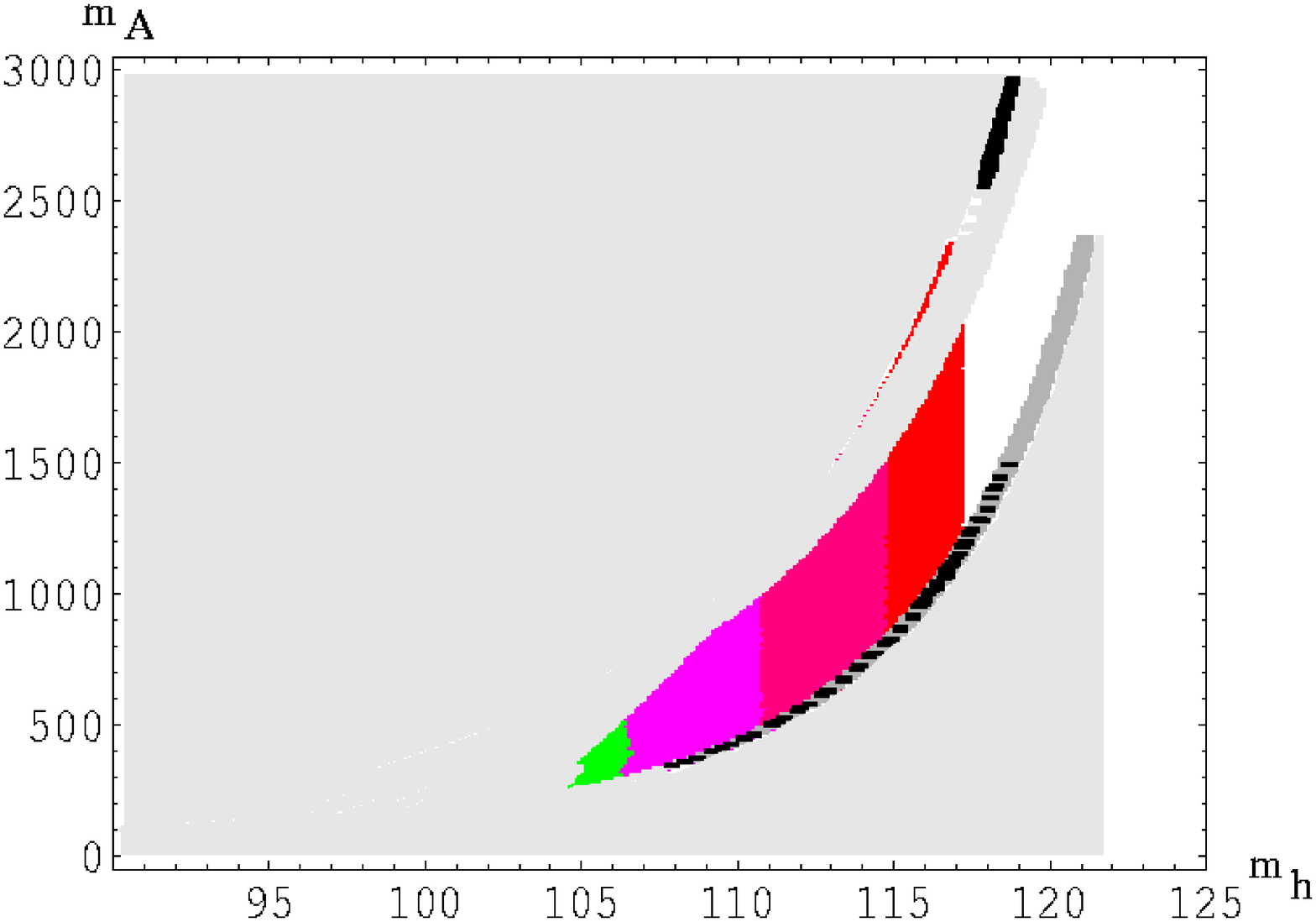,width=7.5cm,height=7.5cm}\hspace*{0.9cm}
      \epsfig{file=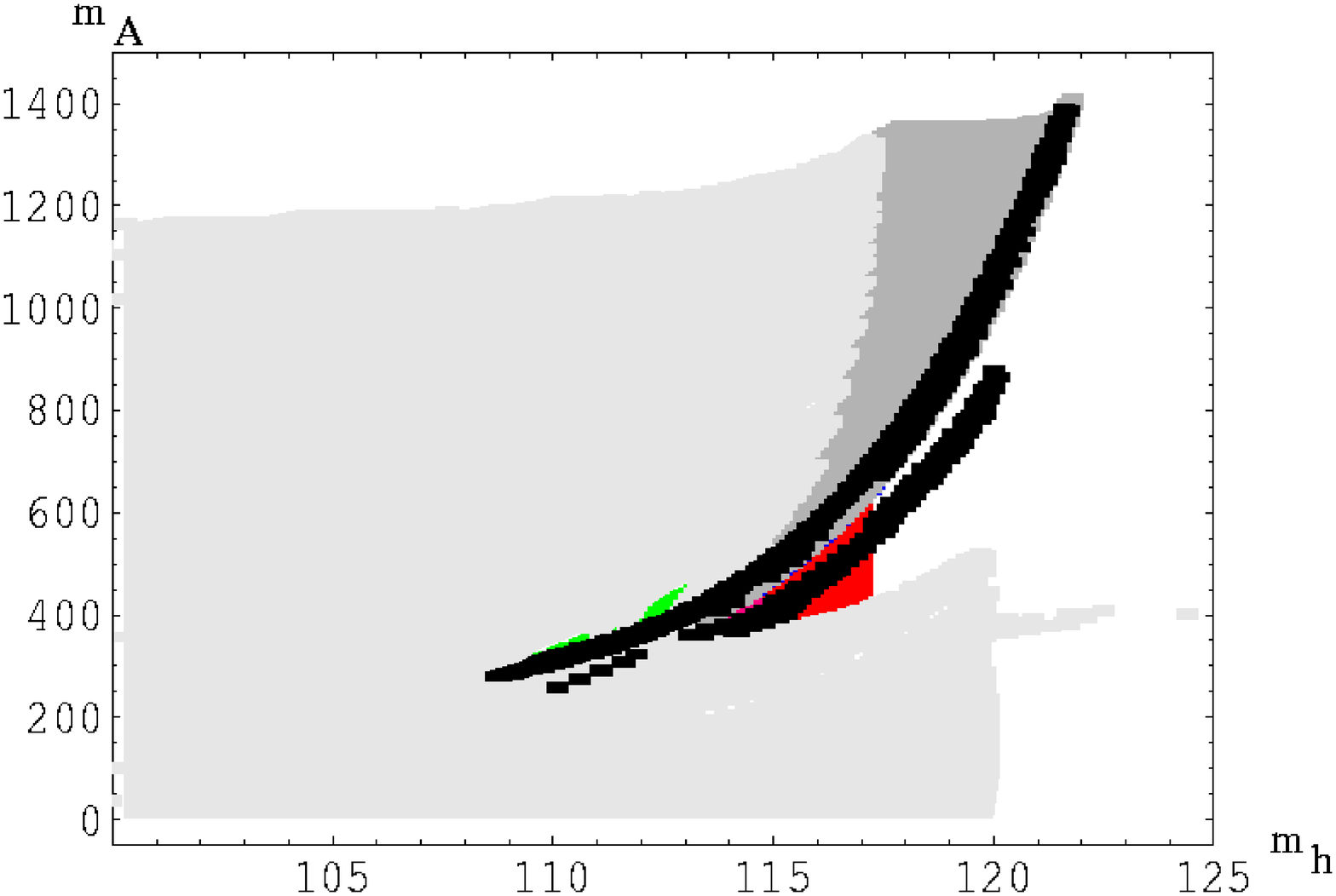,width=7.5cm,height=7.5cm} }
\vspace*{.1cm}
\end{center}
\caption{\it The mSUGRA  ($m_h, m_A$) parameter space with 
all constraints imposed for $A_0=0$, $\mu>0, \tan\beta=10$ (left) and 
$\tan\beta=50$ (right). The top mass is fixed to  $m_t=172.7$ GeV.}
\end{figure}

In the right frame of Fig.~12, i.e. for $\tan\beta=50$, the {\em lower} black
region is the ``focus point'' region, whereas the upper black strip is the
$\tilde \tau_1$ co--annihilation region merged with the $A-$pole region, as
discussed in the context of Fig.~11. For given $m_A$, $m_h$ is maximized if
$m_{1/2}$ is minimized, i.e. in the ``focus point'' region. The reason is that
$m_A$ is sensitive to $m_{1/2}$ already at the tree--level, through the
relation that fixes $\mu^2$ in terms of $M_Z^2$ and the soft breaking
parameters; in contrast, if $m_A^2 \gg M_Z^2$, $m_h$ depends on $m_{1/2}$ only
through loop effects. We also note that the accessible region of the $(m_h,
m_A)$ plane becomes very narrow for large $\tan\beta$.


\section*{4. Sparticle and Higgs boson mass bounds}

The figures shown in Sec.~3.1 show that the allowed region in the $(m_0,
m_{1/2})$ plane depends very strongly on the value of $\tan\beta$.  There is
also a significant dependence on $A_0$. Finally, even though the top mass is
now the (relatively) most accurately known of all quark masses, we saw that
varying $m_t$ within its current limits still moves the boundaries of allowed
regions by hundreds or, in case of the ``focus point'' region, even thousands
of GeV. Similar shifts of the allowed region occur when plotted in terms of
pairs of physical masses, as shown in Sec.~3.2.

One simple way to show the {\em total} allowed ranges of physical masses is to
scan over the entire parameter space that is consistent with a given set of
constraints; this is the topic of this Section. We saw in Sec.~2 that not all
constraints should be treated on an equal footing. Briefly, lower bounds on
masses (or cross sections or branching ratios) from accelerator--based
experiments are most robust, since both beam and detector are (hopefully) well 
controlled by the experimenter. Bounds on masses, and on cross sections of
processes that can occur at tree level, usually also do not have many
theoretical ambiguities. In contrast, we saw that one can evade the $b
\rightarrow s \gamma$ constraint by a relatively minor modification of the
model \cite{leszek}. In the case of the $g_\mu-2$, there is the additional 
ambiguity due to the $\sim 2\sigma$ discrepancy between SM `predictions' based 
on different data sets, see eqs.~(\ref{gmuc1}) and (\ref{gmuc2}). Finally, the 
DM constraint (\ref{dm}) required several (reasonable) assumptions for its
derivation, and needs additional assumptions to be translated into allowed
regions in the mSUGRA parameter space.

It was originally hoped that the upper bound on the DM relic density (the
so--called `overclosure' constraint) would allow to establish reliable, useful
upper bounds on sparticle masses. Under the standard assumptions listed in
Sec.~2, the predicted LSP relic density is proportional to the inverse of the
(effective) LSP annihilation cross section into SM particles (or MSSM Higgs
bosons, if kinematically accessible). This cross section in turn (through
dimensional arguments, or by unitarity) scales like the inverse square of the
relevant mass scale. Indeed, unitarity does allow to establish an upper bound
on the mass of any WIMP; however, this bound exceeds 100 TeV \cite{unit}, and
is thus not particularly useful, since we lack the means to build colliders
that could cover this kind of mass range. In the context of mSUGRA, it became
clear quite early on \cite{dn3} that very, even ``unnaturally'', large
masses can be compatible with the DM constraint (\ref{dm}) even in standard
cosmology.

On the other hand, we did see in Figs.~1--6 that this constraint excludes
large chunks of otherwise allowed parameter space. One might therefore think
that it would at least affect the lower bounds on sparticle masses
significantly. Table~1 shows that this is not really the case. This table
lists lower bounds on the masses of some new (s)particles, imposing various
sets of constraints. We always impose all constraints discussed up to and
including eq.~(\ref{gmuc1}) in Sec.~2. Sets I through III and IV through VI
differ in that they are based on the conservative $g_\mu - 2$ constraint
(\ref{gmuc1}) and the more aggressive constraint (\ref{gmuc2}), respectively.
Since the latter requires a positive supersymmetric contribution to $g_\mu$,
it allows us to derive {\em upper} bounds as well as lower bounds on the
masses of new (s)particles. In addition to these basic constraints, sets II
and V impose the $b \rightarrow s \gamma$ constraint (\ref{bsg}), and sets III
and VI in addition impose the DM constraint (\ref{dm}).

\begin{table}[h!]
\begin{center}
\caption{\it Lower bounds on the masses of superparticles and Higgs bosons, and
  upper bound on the LSP--nucleon scattering cross section, derived in mSUGRA
  under six different sets of assumptions. ``HWIP'' and ``HSIP'' stand for
  ``heaviest weakly interacting particle'' and ``heaviest strongly interacting
  particle'', respectively. In all cases the constraints
  discussed in Sec.~2 up to the requirement (\ref{gmuc1}) are imposed. In
  Set~II we in addition impose the constraint (\ref{bsg}) from $b \rightarrow
  s \gamma$ decays (including the sign of the decay amplitude). Including in
  addition the DM constraint (\ref{dm}) leads to Set~III. Sets IV--VI are like
  Sets I--III, except that the more conservative $g_{\mu} - 2$ constraint
  (\ref{gmuc1}) has been replaced by the more aggressive requirement
  (\ref{gmuc2}); in these cases we give allowed ranges, rather than only
  one--sided bounds. (Upper bounds on masses in Set~V are the same as in
  Set~IV.) All limits have been obtained by scanning the parameter space
  (\ref{paras}), for 166.9 GeV $\leq m_t \leq$ 178.5 GeV.}
\vspace*{5mm}
\begin{tabular}{|c||c|c|c||c|c|c|}
\hline
Quantity & Set I & Set II & Set III & Set IV & Set V & Set VI \\
\hline 
$m_{\tilde e_R} \simeq m_{\tilde \mu_R}$ [GeV] & 106 & 106 & 107 & [106,1320]
 & 106 & [108,1300] \\
$m_{\tilde e_L} \simeq m_{\tilde \mu_L}$ [GeV] & 152 & 168 & 169 & [152,1330]
 & 168 & [171,1310] \\
$m_{\tilde \tau_1}$ [GeV] & 99 & 99 & 99 & [99,1020] & 99 & [99,915] \\
$m_{\tilde \tau_2}$ [GeV] & 156 & 171 & 172 & [156,1160] & 171 & [174,1130] \\
$m_{\tilde \nu_\tau}$ [GeV] & 130 & 149 & 149 & [130,1160] & 149 & [152,1120]\\
\hline
$m_{\tilde \chi_1^\pm}$ [GeV] & 105 & 105 & 105 & [105,674] & 105 & [105,667]\\
$m_{\tilde \chi_2^\pm}$ [GeV] & 218 & 218 & 233 & [219,1003] & 227 & [337,999]
 \\
$m_{\tilde \chi_1^0}$ [GeV] & 52 & 52 & 53 & [52,359] & 53 & [55,357] \\
$m_{\tilde \chi_2^0}$ [GeV] & 105 & 105 & 105 & [105,674] & 105 & [106,667] \\
$m_{\tilde \chi_3^0}$ [GeV] & 135 & 135 & 135 & [135,996] & 135 & [292,991] \\
$m_{\tilde \chi_4^0}$ [GeV] & 217 & 218 & 234 & [218,1003] & 226 & [337,999] \\
\hline
$m_{\tilde g}$ [GeV] & 359 & 380 & 380 & [361,1880] & 399 & [412,1870] \\
$m_{\tilde d_R} \simeq m_{\tilde s_R}$ [GeV] & 406 & 498 & 498 & [406,1740]
 & 498 & [498,1740] \\
$m_{\tilde d_L} \simeq m_{\tilde s_L}$ [GeV] & 424 & 518 & 518 & [424,1810]
 & 518 & [518,1800] \\
$m_{\tilde b_1}$ [GeV] & 294 & 459 & 463 & [295,1520] & 459 & [463,1500] \\
$m_{\tilde b_2}$ [GeV] & 400 & 498 & 498 & [400,1600] & 498 & [498,1590] \\
$m_{\tilde t_1}$ [GeV] & 102 & 104 & 104 & [102,1440] & 231 & [244,1440] \\
$m_{\tilde t_2}$ [GeV] & 429 & 547 & 547 & [431,1600] & 547 & [547,1590] \\
\hline
$m_h$ [GeV] & 91 & 91 & 91 &  [91,124] & 91 & [91,124] \\
$m_H$ [GeV] & 111 & 111 & 111 & [111,975] & 111 & [111,954] \\
$m_{H^\pm}$ [GeV] & 128 & 128 & 128 & [128,979] & 128 & [128,960] \\
\hline
$m_{\rm HWIP}$ [GeV] & 349 & 362 & 366 & [351,1330] & 366 & [371,1310] \\
$m_{\rm HSIP}$ [GeV] & 432 & 556 & 556 & [432,1880] & 556 & [566,1870] \\
\hline
$\sigma_{\tilde \chi_1^0 p}$ [ab] & 140 & 140 & 7.5 & [$10^{-4}$,140] & 140 
& [$10^{-4}$,7.5] \\
\hline
\end{tabular}
\end{center}
\end{table}

We see that the lower bounds on the masses of some key (s)particles always
saturate their current bounds from collider physics, no matter what additional
constraints we impose. This is true, in particular, for the lighter chargino,
the lightest charged slepton (always $\tilde \tau_1$ in mSUGRA), as well as
both the CP--even Higgs bosons. The lower bound on the masses of the charged
Higgs boson is also independent of the constraints imposed; it follows almost
directly from the structure of the MSSM. The lower bounds on the masses of the
gluino and the lighter two neutralinos are to a large extent fixed by
gaugino mass unification and the chargino mass bound. In particular, the bound
on $m_{\tilde \chi_2^0}$ is practically identical to that on $m_{\tilde
  \chi_1^\pm}$. Note that gaugino mass unification holds for running
($\overline{\rm DR}$) masses. Going to the pole mass entails potentially quite
substantial radiative corrections in case of the gluino \cite{martinvaughn}.
The lower bound on $m_{\tilde g}$ therefore increases by about 15\% when going
from the loosest constraints (Set~I) to the tightest ones (Set~VI), the
biggest increase being due to the $b \rightarrow s \gamma$ constraint, which
excludes scenarios with large $\tan\beta$ and relatively small squark masses.

The same effect is also visible in the lower bounds on first and second
generation squark masses themselves, which increase by about 20\% when this
constraint is imposed. Note that even for the loosest set of constraints,
Set~I, the lower bounds on the masses of first and second generation squarks
are significantly above current search limits \cite{pdg}. This is a
consequence of the assumed universality of scalar masses, together with the
requirement of having sufficiently large soft breaking masses in the stop
sector to satisfy the Higgs search limits. These bounds are therefore
saturated for the largest possible $m_t$ value.

The masses of the heavier neutralinos and charginos also depend only weakly on
the set of constraints imposed. Most of these lower bounds again follow
directly from the structure of the MSSM with gaugino mass unification,
together with the LEP bound on $m_{\tilde \chi_1^\pm}$; the universality of
scalar masses and $A-$parameters, which are defining properties of mSUGRA,
play little role here. The only significant exception is the increase of
$m_{\tilde \chi_3^0}$ for the most restrictive set of constraints (Set~VI).
This bound is saturated in the ``focus point'' region of large $m_0$, where
the supersymmetric contribution to $g_\mu - 2$ tends to be below the range
(\ref{gmuc2}).

The lower bounds on the masses of third generation squarks are the quantities
that are most sensitive to the additional constraints imposed in Sets II
through VI. In particular, requiring both the more aggressive $g_\mu - 2$
constraint (\ref{gmuc2}) and the $b \rightarrow s \gamma$ constraint
(\ref{bsg}) more than doubles the lower bound on $m_{\tilde t_1}$, from the
LEP limit of $\sim 100$ GeV that can be saturated for Sets I through IV, to
about 240 GeV. Combinations of parameters leading to a light $\tilde t_1$
which are compatible with the $b \rightarrow s \gamma$ constraint have
relatively small $\tan\beta$, but large $m_0$; this leads to a supersymmetric
contribution to $g_\mu-2$ below the range (\ref{gmuc2}). One can also generate
a light $\tilde t_1$ by taking modest values of $m_0$ and $m_{1/2}$, in
agreement with the aggressive $g_\mu - 2$ constraint (\ref{gmuc2}); however,
this requires very large values of $|A_0| / m_0$, which leads to a violation
of the $b \rightarrow s \gamma$ constraint (\ref{bsg}). This latter constraint
by itself also increases the lower bound on $m_{\tilde b_1}$ by about 50\%,
since a light $\tilde b_1$ requires large $\tan\beta$ (which maximizes the
bottom Yukawa coupling, as well as $\tilde b_L - \tilde b_R$ mixing), which in
turn leads to large (negative, for $\mu > 0$) supersymmetric contributions to
$b \rightarrow s \gamma$ decays.

In contrast, imposing in addition the DM constraint (\ref{dm}) has very little
effect on the lower bounds on sparticle masses. It does, however, drastically
reduce the maximal possible elastic spin--independent LSP--proton scattering
cross section, which is shown in the last line of the Table. The calculation
of this cross section is based on refs.~\cite{dn4}. In mSUGRA the potentially
largest contributions come from the exchange of the heavier CP--even Higgs
boson. Since increasing $\tan\beta$ both reduces the mass of this boson and
increases its coupling to down--type quarks, the cross section grows quickly
when $\tan\beta$ becomes larger. In addition, it is maximized by increasing
the coupling of the LSP to Higgs bosons, which requires rather larger
gaugino--higgsino mixing; this cross section is therefore largest in the
``focus point'' region \cite{sig_foc}. However, the same coupling also leads
to too effective LSP annihilation, resulting in too low a relic density.
Imposing the lower bound on the relic density in (\ref{dm}) therefore reduces
the upper bound on this cross section by about a factor of 20. We should
mention here that even the reduced value of 7.5 ab, which is saturated for an
LSP mass near 160 GeV, exceeds the current experimental lower limit on this
quantity, if standard assumptions about the distribution of the LSPs in the
halo of our galaxy are correct \cite{cdms}.  Due to the uncertainties in this
distribution we have not included LSP search limits in our set of constraints.

While all sets of constraints allow the masses of some weakly interacting
sparticles to lie right at the current experimental limit, mSUGRA implies that
the {\em heaviest} weakly interacting new particle (sparticle or Higgs boson)
must lie above $\sim 350$ GeV at least. Note that this limit lies well above
the lower bound on the mass of any one weakly interacting (s)particle. The
reason is that these bounds cannot be saturated simultaneously. For example,
the lower bounds on slepton masses are saturated at moderate values of
$\tan\beta$, big enough to avoid excessive lower bounds from Higgs searches,
but not so big as to imply strong lower bounds from the $g_\mu- 2$ constraint
(\ref{gmuc1}). In contrast, the lower bounds on the masses of the heavy Higgs
bosons are saturated at very large values of $\tan\beta$. Similarly, the lower
bound on the mass of the heaviest strongly interacting sparticle in a given
spectrum is somewhat larger than the largest lower bound considered
separately.

As mentioned earlier, imposing the aggressive $g_\mu - 2$ constraint
(\ref{gmuc2}) allows to derive {\em upper} bounds in the masses of sparticles
and Higgs bosons. The reason is that the supersymmetric contribution, which
comes from gaugino--slepton loops, would vanish if either the gaugino masses or
the slepton masses became very large. This leads to upper bounds on both $m_0$
and $m_{1/2}$, as can be seen by studying the blue regions in Figs.~1 through
6. This in turn imposes an upper bound on $|\mu|$ via the condition of
radiative electroweak symmetry breaking. As a result, {\em all} sparticle and
Higgs masses can be bounded in mSUGRA using this single constraint! We
emphasize that one needs to assume universality of both scalar and gaugino
masses to derive these constraints. Numerically, the upper bounds on the
masses of the first generation squarks as well as on the gluino mass imply
that discovery of supersymmetry at the LHC should be straightforward
\cite{lhc}. Unfortunately even in this favorable scenario discovery of
charginos would not be guaranteed even at a 1 TeV $e^+e^-$ collider, and
discovery of sleptons would be guaranteed only at a CLIC--like machine
operating at $\sqrt{s} \gsim 3$ TeV. On the other hand, imposing this
constraint reduces the upper bound on $m_h$ to 124 GeV (which becomes 127 GeV
if one allows for a 3 GeV theoretical uncertainty), which might perhaps even
give the Tevatron a chance to probe this scenario (with probably less than
compelling statistical significance, alas) \cite{tevhiggs}. Imposing in
addition the $b \rightarrow s \gamma$ constraint does not change these upper
bounds at all; even imposing the DM constraint (\ref{dm}) leaves these upper
bounds almost unchanged.

One lesson from Table~1 is that imposing the DM constraint (\ref{dm}) has
little effect on either upper or lower bounds on the masses of sparticles and
Higgs bosons in mSUGRA, {\em if} one scans over the entire allowed parameter
space. Figs.~13 show that the dramatic reduction of the allowed region that
results from this constraint that is evident in the Figures presented in
Sec.~3 does narrow down the possible ranges of masses when $\tan\beta$ is kept
fixed. These figures compare the allowed ranges of the masses of $\tilde e_R,
\, \tilde \chi_1^\pm, \, \tilde t_1$ and $H$ for the constraint Sets IV and VI
of Table~1. In the former case one can saturate the LEP limit on the mass of
the lighter chargino for any $\tan\beta \geq 5$. However, light charginos are
DM--compatible only in the ``focus point'' region, which in turn is (barely)
compatible with the $g_\mu - 2$ constraint (\ref{gmuc2}) only at large
$\tan\beta$. Moreover, at very large $\tan\beta$ the $b \rightarrow s \gamma$
constraint becomes quite severe. As a result, constraint Set~VI allows to
saturate the LEP chargino mass bound only for $40 \lsim \tan\beta \lsim 50$.

\begin{figure}[h!]
\begin{center}
\vspace*{-.5cm}
\rotatebox{-90}{\includegraphics[width=0.62\textwidth]{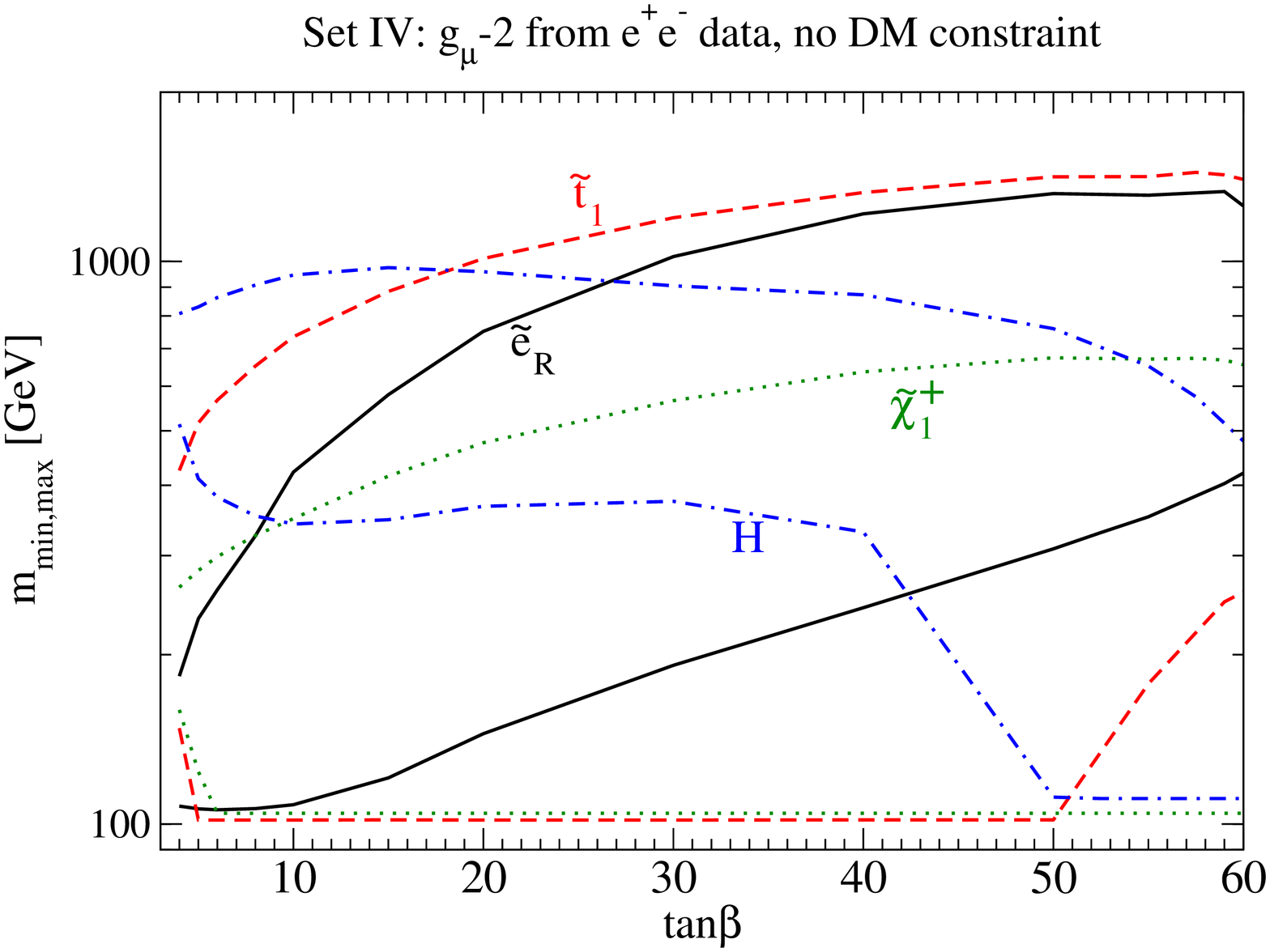}}

\vspace*{-.5cm} \noindent
\rotatebox{-90}{\includegraphics[width=0.62\textwidth]{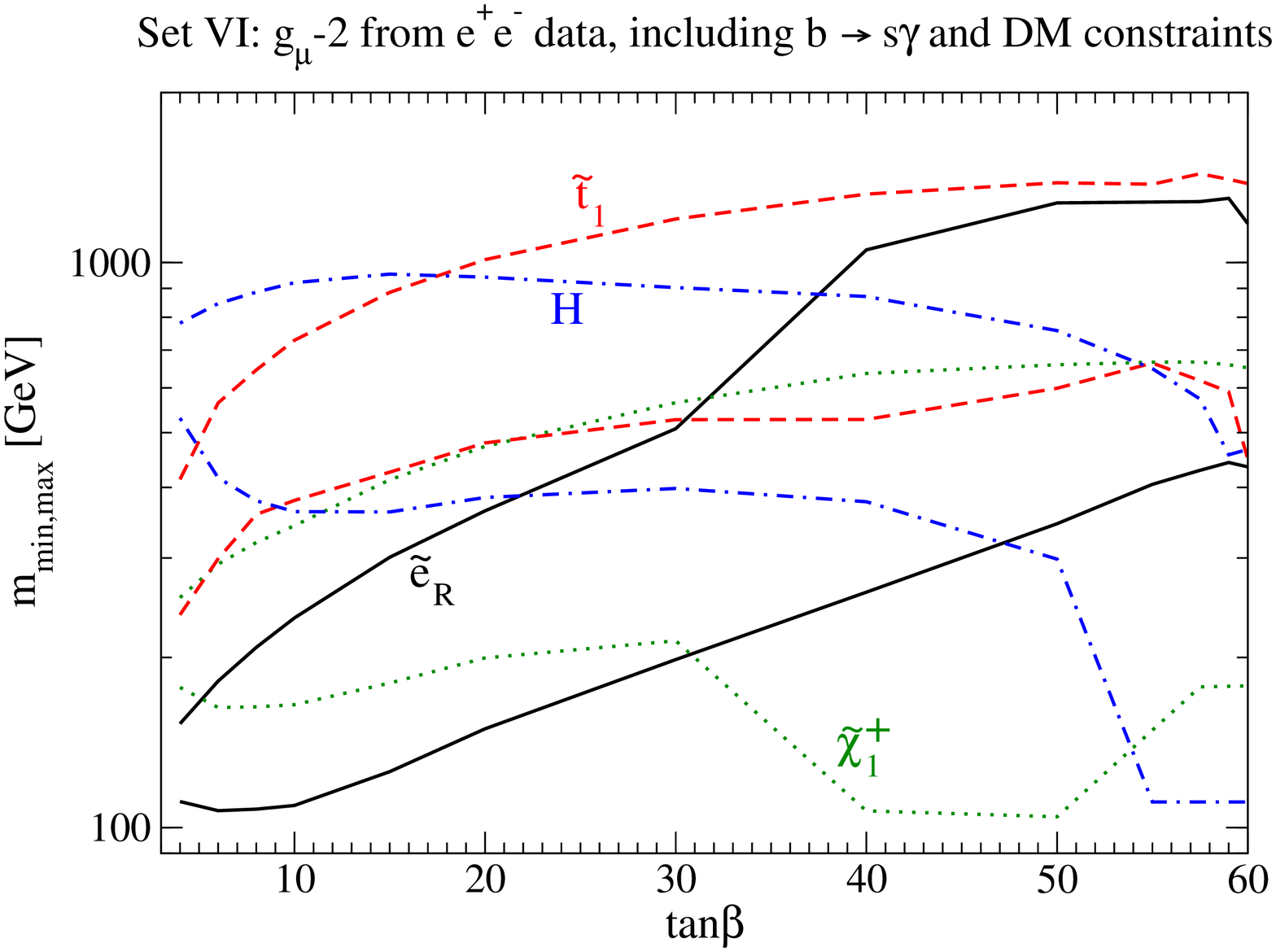}}
\vspace*{.1cm}
\caption{\it The minimal and maximal value of select sparticle and Higgs
  masses: solid (black) curves: $\tilde e_R$; dashed (red) curves: $\tilde
  t_1$; dotted (green) curves: $\tilde \chi_1^\pm$; dot--dashed (blue) curves:
  $H$. The upper (lower) figure shows the bounds without (with) imposing the
  DM and $b\to s\gamma$ constraints.} 
\end{center}
\vspace*{-7mm}
\end{figure}

The combined effect of the DM and $b \rightarrow s \gamma$ constraints on the
lower bound on $m_{\tilde t_1}$ is even more dramatic. Without these
constraints, the LEP limit on this mass can be saturated for any $\tan\beta
\leq 50$. However, as we already saw in Table~1, the $b \rightarrow s \gamma$
constraint increases the lower bound on this mass bound by more than a factor
of 2, if one insists on the aggressive $g_\mu - 2$ constraint (\ref{gmuc2});
Fig.~13 shows that this lower bound then also increases quite rapidly with
increasing $\tan\beta$. As a result, if $\tan\beta$ was known, imposing
constraint Set~VI would allow to predict $m_{\tilde t_1}$ to within a factor
of $\sim 3$. However, since the allowed band moves upward with increasing
$\tan\beta$, we can currently predict $m_{\tilde t_1}$ only within a factor of
$\sim 6$, even if we impose this most restrictive set of constraints, as shown
in Table~1.

Imposing the DM constraint (\ref{dm}) reduces the upper bounds on sparticle
masses for fixed $\tan\beta$. This effect is quite mild in most cases, but
becomes prominent for $\tilde e_R$ at small and moderate values of
$\tan\beta$. The $g_\mu - 2$ constraint (\ref{gmuc2}) by itself already leads
to a strong $\tan\beta$ dependence of this bound; recall that the
supersymmetric contribution to this quantity is essentially proportional to
$\tan\beta$. We saw at the end of Sec.~3.1 that for small and moderate
$\tan\beta$ the only overlap of the DM and (aggressive) $g_\mu-2$ allowed
regions occurs in the $\tilde \tau$ co--annihilation region, which has
relatively large gaugino masses; as a result, one needs even smaller slepton
masses to produce a sufficiently large $g_\mu-2$. However, once $\tan\beta
\geq 40$, one can satisfy this last constraint even in the ``focus point''
region; imposing the DM constraint then has little effect on the upper bound
on $m_{\tilde e_R}$.  Finally, we note that the allowed range of $m_H$ is
fixed almost completely by the ``base'' set of constraints plus the aggressive
$g_\mu - 2$ constraint (\ref{gmuc2}); imposing in addition the $b \rightarrow
s \gamma$ and DM constraints has little effect here.

\section*{5. Summary and Conclusions}

In this paper we provide an updated scan of the mSUGRA parameter space. This
includes the new central value of the mass of the top quark, the inclusion of
additional higher order corrections to the mass of the lightest CP--even Higgs
boson $h$, and new information on the sign of the matrix element for $b
\rightarrow s \gamma$ decays from the analysis of $b \rightarrow s \ell^+
\ell^-$ decays. 

We find that the reduction of the central value of $m_t$ from 178 GeV to about
173 GeV shifts the allowed parameter space significantly. This is the
consequence of two effects:

\begin{itemize}
  
\item The corrections to $m_h^2$ scale like the fourth power of $m_t$, but
  only scale logarithmically with the sparticle (mostly stop) mass scale. As a
  result, a few percent reduction of $m_t$ has to be compensated by an
  increase of $m_{\tilde t}$ of up to several tens of percent. This relative
  shift increases with $m_{\tilde t}$, and is therefore most prominent for
  smaller $\tan\beta$, where the LEP Higgs search constraint is most severe.
  
\item The location of the ``focus point'' region at $m_0^2 \gg m_{1/2}^2$
  where the LSP acquires a significant higgsino component depends very
  sensitively on $m_t$. The higgsino component of $\tilde \chi_1^0$ is only
  sizable if $|\mu|$ is rather small. This will happen if the squared soft
  breaking mass of the Higgs boson coupling to the top quark is small or
  positive at the ``weak'' scale $Q_W$, which should be chosen close to
  $\sqrt{ m_{\tilde t_1} m_{\tilde t_2}}$ to get good convergence of the
  perturbative series. This parameter in turn can be written as
\beq \label{mhsq}
m^2_{H_u}(Q_W) = a m_0^2 + b m_{1/2}^2 + c A_0^2 + d A_0 m_{1/2},
\eeq
where the dimensionless coefficients $a,b,c,d$ depend on the dimensionless
MSSM couplings as well as (logarithmically) on $Q_W$. The crucial observation
is that $|a| \ll 1$ for $m_t \sim 175$ GeV and not too small $\tan\beta$
(recall that the top Yukawa coupling at the weak scale is $\propto
1/\sin\beta$). In contrast, $|b|$ is quite sizable, with $b < 0$. A small
$m^2_{H_u}(Q_W)$ is therefore only possible if $m_0^2 \gg m_{1/2}^2$ {\em and}
$a \geq 0$. Since $a$ depends (roughly) quadratically on $m_t / \sin\beta$,
but only logarithmically on the sparticle mass scale (through $Q_W$), a small
change of $m_t$ therefore leads to a large shift of the value of $Q_W$, or,
equivalently, of $m_0$ where the ``focus point'' region starts. 

\end{itemize}

The first effect makes it more difficult to reconcile, at low values of
$\tan\beta$, the Higgs mass constraint from LEP with the evidence for a
positive contribution to the anomalous magnetic moment of the muon. However,
for larger $\tan\beta$ the Higgs mass bound allows smaller sparticle masses,
while the contribution to $g_\mu-2$ remains significant for larger sparticle
masses. As a result, even for $m_t = 173$ GeV both constraints can be
satisfied simultaneously for $\tan\beta \gsim 10$.

By far the most stringent constraint on the parameter space comes from the
requirement that the lightest neutralino should have the correct relic
density. As discussed in Sec.~2, this constraint can only be translated into
bounds on the mSUGRA parameter space if several assumptions are made. Under
the usual assumption of thermal LSP production and standard cosmology, only a
few discrete ``DM allowed'' regions survive. Out of these, the ``bulk'' region
of moderate $m_0$ and moderate $m_{1/2}$ is affected most by the reduced mass
of the top quark; in fact, it disappears altogether if $m_t$ is indeed near
its current central value of $\sim 173$ GeV. The $\tilde \tau_1$
co--annihilation region is also reduced in size, since the region excluded by
Higgs searches now extends to larger values of $m_{1/2}$. On the other hand,
as noted above, the region where the LSP has significant higgsino component
becomes {\em larger} when $m_t$ is reduced. Similarly, the lowest value of
$\tan\beta$ where $2 m_{\tilde \chi_1^0} \simeq m_A$ (the $A-$pole region)
becomes smaller; this region also becomes broader, thanks to the increased
strength of the $A \tilde \chi_1^0 \tilde \chi_1^0$ coupling. In contrast, the
conceptually similar $h-$pole region is much reduced in size.

We also provided views of the mSUGRA parameter space plotted in the plane
spanned by two physical sparticle or Higgs boson masses. Whereas some masses
are essentially independent of each other (e.g. $m_{\tilde e_R}$ and
$m_{\tilde \chi_1^0}$), others are strongly correlated (e.g. $m_A$ and $m_h$);
most pairs are intermediate between these extremes.

In addition to these scans of parameter space, we also provided upper and
lower {\em bounds} on the masses of Higgs bosons and sparticles. Here it is
crucial to properly include the uncertainty of the input parameters, in
particular, of $m_t$. This sensitivity comes mostly through the dimensionless
coefficients in eq.(\ref{mhsq}), as well as the analogous expression for
$m^2_{H_d}$. These coefficients determine $|\mu|$ through the condition of
electroweak symmetry breaking, which affects the spectra of neutralinos,
charginos and Higgs bosons. The latter is also directly dependent on these
coefficients; through the Higgs search limits, they then affect (the lower
bounds on) all sparticle masses. It should be noted that these coefficients
also depend on other input parameters, in particular on $\alpha_S$ and
$m_b$. However, this dependence is much milder than that on $m_t$. We
therefore believe that varying $m_t$ over its entire currently allowed
$2\sigma$ range, while keeping $\alpha_S$ and $m_b$ fixed to their  default
values, gives a reasonable estimate of the effect of the current input
parameter uncertainties.

The somewhat surprising result of these scans is that the masses of many
superparticles and Higgs bosons can still lie right at their current limits
from collider searches even if the most restrictive set of constraints is
applied, including the Dark Matter constraint and the more aggressive
interpretation of the $g_\mu - 2$ constraint. This means that ongoing and
near--future experiments still have good chances to discover new particles
even in this very constrained version of the MSSM. Not all these lower bounds
can be saturated simultaneously, however. As a result, the most robust
constraints (essentially the collider limits plus a conservative version of
the $g_\mu-2$ constraint, with no DM requirement) by themselves already imply
that a 500 GeV linear collider will not be able to discover all new weakly
interacting (s)particles; one will need an energy of at least $\sim 700$ GeV
to achieve this. We also saw that these lower bounds are in most cases
surprisingly insensitive to the introduction of new constraints; in
particular, requiring the lightest neutralino to be a good thermal DM
candidate does not shift them much.

Useful upper bounds on the masses of sparticles and Higgs bosons (with the
exception of the lightest CP--even Higgs boson, see ref.~\cite{newhiggs}) can
only be derived if we assume that a positive supersymmetric contribution to
$g_\mu - 2$ is indeed required, as is indicated (at the $\sim 2.5\sigma$
level) when data from $e^+ e^- \rightarrow {\rm hadrons}$ are used to
calculate the SM prediction for this quantity. This imposes upper bounds on
the masses of both sleptons and gauginos; in the mSUGRA context this implies
upper bounds on both $m_0$ and $m_{1/2}$, which leads to upper bounds on {\em
  all} new (s)particles. Quantitatively, we find that this requirement by
itself implies that strongly interacting sparticles must be within the reach
of the LHC.  Moreover, a $\sim 1$ TeV $e^+e^-$ collider would then be able to
at least discover superparticles, in the $\tilde \chi_1^0 \tilde \chi_2^0$
channel.  However, even in this case one may need a CLIC--like collider, with
center of mass energy nearly reaching 3 TeV, to discover all new weakly
interacting (s)particles. We stress again that these (upper) bounds have been
obtained by scanning over the entire parameter space, including scanning over
the $2\sigma$ range of $m_t$. Imposing in addition the Dark Matter constraint
narrows down the allowed ranges of some masses if $\tan\beta$ is held fixed,
but has little effect on the absolute upper bounds after scanning over the
entire allowed parameter space.

We conclude that mSUGRA remains viable. In fact, even after imposing all
plausible experimental and theoretical constraints the allowed parameter space
still contains a large variety of different spectra, with quite different
phenomenology. Even if supersymmetry provides the missing Dark Matter in the
Universe and explains the possible excess of the anomalous dipole moment of
the muon, superparticles might be out of reach of both the Tevatron and the
first stage of a linear $e^+e^-$ collider; on the other hand, it is also
entirely possible that their discovery is ``just around the corner''.

\subsubsection*{Acknowledgments}
MD thanks the Center of Theoretical Physics at Seoul National University, as
well as the School of Physics at KIAS, for hospitality while this project was
nearing completion.



\begin{thebibliography}{199} 
%
\bibitem{mSUGRA} 
A.H. Chamseddine, R. Arnowitt and P. Nath, Phys. Rev. Lett. {\bf 49} (1982)
970; 
R. Barbieri, S. Ferrara and C.A Savoy, Phys. Lett. {\bf B119} (1982) 343;
L. Hall, J. Lykken and S. Weinberg, Phys. Rev. {\bf D27} (1983) 2359;
E. Cremmer, P. Fayet and L. Girardello, Phys. Lett. {\bf B122}, 41 (1983);
N. Ohta, Prog. Theor. Phys. {\bf 70}, 542 (1983).
%
\bibitem{nilles}
H.P. Nilles, Phys. Rep. {\bf 110} (1984) 1.
%
\bibitem{hierarchy}
E. Witten, Nucl. Phys. {\bf B188}, 513 (1981);
N. Sakai, Z. Phys. {\bf C11}, 153 (1981);
S. Dimopoulos and H. Georgi, Nucl. Phys. {\bf B193}, 150 (1981);
R.K. Kaul and P. Majumdar, Nucl. Phys. {\bf B199}, 36 (1982).
%
\bibitem{GaugeUni} 
J. Ellis , S. Kelley and D.V. Nanopoulos, Phys. Lett. 
{\bf B260} (1991) 131; 
U. Amaldi, W. de Boer and H. F\"urstenau, Phys. Lett. {\bf B260} (1991) 447; 
P. Langacker and M. Luo, Phys. Rev. {\bf D44} (1991) 817;
C. Giunti, C.W. Kim and U.W. Lee, Mod. Phys. Lett. {\bf A6} (1991) 1745.
%
\bibitem{olddm}
H. Goldberg, Phys. Rev. Lett. {\bf 50}, 1419 (1983);
J. Ellis, J. Hagelin, D.V. Nanopoulos, K. Olive and M. Srednicki, Nucl. Phys.
{\bf B238}, 453 (1984).
%
\bibitem{dmrev}
For a review, see G. Jungman, M. Kamionkowski and K. Griest,
Phys. Rep. {\bf 267} (1996) 195, hep-ph/9506380.
%
\bibitem{radbreak}
L.E. Ib\'a\~nez and G.G. Ross, Phys. Lett. {\bf 110B}, 215 (1982); 
L.E. Ib\'a\~nez, Phys. Lett. {\bf 118B}, 73 (1982); 
J. Ellis, D.V. Nanopoulos and K. Tamvakis, Phys. Lett. {\bf 121B}, 123 (1983); 
L. Alvarez-Gaum\'e, J. Polchinski and M.B. Wise, Nucl. Phys. {\bf B221}, 495 
(1983).
%
\bibitem{guts}
D. Auto, H. Baer, C. Balazs, A. Belyaev, J. Ferrandis and X. Tata, JHEP {\bf
  0306}, 023 (2003), hep-ph/0302155; 
R. Dermisek, S. Raby, L. Roszkowski and R. Ruiz De Austri, JHEP {\bf 0304},
037 (2003), hep--ph/0304101; 
M.R. Ramage, Nucl. Phys. {\bf B720}, 137 (2005), hep--ph/0412153. 
%
\bibitem{nonuniv}
J.R. Ellis, T. Falk, K.A. Olive and Y. Santoso, Nucl. Phys. {\bf B652}, 259
(2003), hep--ph/0210205; 
D. Auto, H. Baer, A. Belyaev and T. Krupovnickas, JHEP {\bf 0410}, 066 (2004), 
hep--ph/0407165; 
H. Baer, A. Belyaev, T. Krupovnickas and A. Mustafayev, JHEP {\bf 0406}, 044
(2004), hep--ph/0403214; 
H. Baer, A. Mustafayev, S. Profumo, A. Belyaev and X. Tata, Phys. Rev. {\bf
  D71}, 095008 (2005), hep--ph/0412059;
G. B\'elanger, F. Boudjema, A. Cottrant, A. Pukhov and A. Semenov,
Nucl. Phys. {\bf B706}, 411 (2005), hep--ph/0407218, and
hep--ph/0412309;  
J.R. Ellis, K.A. Olive, Y. Santoso and V.C. Spanos, Phys. Lett. {\bf B603}, 51
(2004), hep--ph/0408118;
H. Baer, A. Mustafayev, E.-K. Park and S. Profumo, JHEP {\bf 0507}, 046 (2005),
hep--ph/0505227;       
Y. Mambrini and E. Nezri, hep--ph/0507263; 
H. Baer, T. Krupovnickas, A. Mustafayev, E.-K. Park, S. Profumo and X. Tata,
hep--ph/0511034.   
%
\bibitem{rela}
J.R. Ellis, K.A. Olive, Y. Santoso and V.C. Spanos, Phys. Lett. {\bf B573},
162 (2003), hep--ph/0305212, and Phys. Rev. {\bf D70}, 055005 (2004),
hep--ph/0405110.
%
\bibitem{wmap}
WMAP Collab., D.N. Spergel et al., Astrophys. J. Suppl. {\bf 148}, 175 (2003),
astro--ph/0302209.
%
\bibitem{others}
H. Baer, C. Balazs, A. Belyaev, J.K. Mizukoshi, X. Tata and Y. Wang, JHEP {\bf
  0207}, 050 (2002), hep--ph/0205325;
H. Baer and C. Balazs, JCAP {\bf 0305}, 006 (2003), hep--ph/0303114; 
U. Chattopadhyay, A. Corsetti and P. Nath, Phys. Rev. {\bf D68}, 035005
(2003), hep--ph/0303201;
J.R. Ellis, K.A. Olive, Y. Santoso and V.C. Spanos, Phys. Lett. {\bf B565},
176 (2003), hep--ph/0303043;
M. Battaglia, A. De Roeck, J.R. Ellis, F. Gianotti, K.A. Olive and L. Pape,
Eur. Phys. J. {\bf C33}, 273 (2004), hep--ph/0306219;
R. Arnowitt, B. Dutta and B. Hu, hep--ph/0310103;
J.R. Ellis, K.A. Olive, Y. Santoso and V.C. Spanos, Phys. Rev. {\bf D69},
095004 (2004), hep--ph/0310356; 
M.E. Gomez, T. Ibrahim, P. Nath and S. Skadhauge, Phys. Rev. {\bf D70}, 035014
(2004), hep--ph/0404025 
J.R. Ellis, S. Heinemeyer, K.A. Olive and G. Weiglein, JHEP {\bf 0502}, 013
(2005), hep--ph/0411216. 
%
\bibitem{newhiggs}
B.C. Allanach, A. Djouadi, J.L. Kneur, W. Porod and P. Slavich, JHEP {\bf
  0409}, 044 (2004), hep--ph/0406166, and references therein.
%
\bibitem{newtop}
CDF Collab. and D0 Collab. and The Tevatron Electroweak Working Group,
{\tt hep--ex/0507091}.
%
\bibitem{bsll}
P. Gambino, U. Haisch and M. Misiak, Phys. Rev. Lett. {\bf 94}, 061803 (2005),
hep--ph/0410155.
%
\bibitem{gmutalk}
See e.g. the talk by K. Melnokiv at {\it SUSY2004}, Tsukuba, Japan, June 2004.
%
\bibitem{gmuexp}
Muon $g-2$ Collab., G.W. Bennett et al., Phys. Rev. Lett. {\bf 89}, 101804
(2002), Erratum--ibid. {\bf 89}, 129903 (2002), hep--ex/0208001, and
Phys. Rev. Lett. {\bf 92}, 161802 (2004), hep--ex/0401008.
%
\bibitem{gmuth}
M. Davier, S. Eidelman, A. H\"ocker and Z. Zhang, Eur. Phys. J. {\bf C31}, 503
(2003), hep--ph/0308213;
K. Hagiwara, A.D. Martin, D. Nomura and T. Teubner, Phys. Rev. {\bf D69},
093003 (2004), hep--ph/0312250;
J.F. de Troconiz and F.J. Yndurain, Phys. Rev. {\bf D71}, 073008 (2005),
hep--ph/0402285;
M. Passera, J. Phys. {\bf G31}, R75 (2005), hep--ph/0411168.
%
\bibitem{recent}
B.C. Allanach and C.G. Lester, hep--ph/0507283.
%
\bibitem{nano-higgs}
J. Ellis, D. Nanopoulos, K.A. Olive and Y. Santoso, hep--ph/0509331.
%
\bibitem{dreesmartin}
M. Drees and S.P. Martin, in {\it Electroweak symmetry breaking and new
  physics at the TeV scale}, ed. T.L. Barklow, hep--ph/9504324.
%
\bibitem{ddk1}
A. Djouadi, M. Drees and J.L. Kneur, JHEP {\bf 0108}, 055 (2001),
hep--ph/0107316.
%
\bibitem{suspect} 
A. Djouadi, J.L. Kneur and G. Moultaka, hep--ph/0211331. 
The program can be down--loaded from the web site:
{\tt www.lpta.univ-montp2.fr/\~{}kneur/Suspect}.
%
\bibitem{rge2l}
S.P. Martin and M. Vaughn, Phys. Rev. {\bf D50}, 2282 (1994), hep--ph/9311340;
I. Jack, D.R.T. Jones, S.P. Martin, M. Vaughn and Y. Yamada, Phys. Rev. {\bf
  D50}, 5481 (1994), hep--ph/9407291.
%
\bibitem{oldccb}
J.M. Fr\`ere, D.R.T. Jones and S. Raby, Nucl. Phys. {\bf B222}, 11 (1983);
M. Claudson, L. Hall and I. Hinchliffe, Nucl. Phys. {\bf B228}, 501 (1983).
%
\bibitem{casas} 
J.A. Casas, A. Lleyda and C. Mu\~noz, Nucl.  Phys. {\bf B471} (1996) 3,
hep--ph/9507294. 
%
\bibitem{pdg}
Particle Data Group, S. Eidelman et al, Phys. Lett. {\bf B592}, 1 (2004).
%
\bibitem{gravi}
J.R. Ellis, K.A. Olive, Y. Santoso and V.C. Spanos, Phys. Lett. {\bf B588}, 7
(2004), hep--ph/0312262;
L. Roszkowski, R. Ruiz de Austri and K.-Y. Choi, JHEP {\bf 0508}, 080 (2005),
hep--ph/0408227. 
%
\bibitem{axi}
L. Covi, L. Roszkowski, R. Ruiz de Austri and M. Small, JHEP {\bf 0406}, 003
(2004), hep--ph/0402240.
%
\bibitem{lepsusy}
For an up--to--date summary of sparticle search limits from the LEP
experiments, see {\tt http://lepsusy.web.cern.ch/lepsusy/} .
%
\bibitem{lephiggs}
The ALEPH, DELPHI, L3 and OPAL Collab.s, Phys. Lett. {\bf B565}, 61 (2003), 
hep--ex/0306033.
%
\bibitem{rho}
R. Barbieri and L. Maiani, Nucl. Phys. {\bf B224}, 32 (1983);
C.S. Lim, T. Inami and N. Sakai, Phys. Rev. {\bf D29}, 1488 (1984);
E. Eliasson, Phys. Lett. {\bf 147B}, 65 (1984); 
M. Drees and K. Hagiwara, Phys. Rev. {\bf D42}, 1709 (1990).
%
\bibitem{rho2} 
A.~Djouadi, P.~Gambino, S.~Heinemeyer, W.~Hollik, C.~J\"unger 
and G.~Weiglein, Phys. Rev. Lett. {\bf 78}, 3636 (1997), hep--ph/9612363, and
Phys. Rev. {\bf D57}, 4179 (1998), hep--ph/9710438.
%
\bibitem{gmususy} 
S.P. Martin and J.D. Wells, Phys. Rev. {\bf D64}, 035003 (2001),
hep-ph/0103067. 
%
\bibitem{pepe2}
G. Degrassi and G.F. Giudice, Phys. Rev. {\bf D58}, 053007 (1998),
hep--ph/9803384.
%
\bibitem{bsgsm}
A.L. Kagan and M. Neubert, Eur. Phys. J. {\bf C7}, 5 (1999);
P. Gambino and U. Haisch , JHEP {\bf 0110}, 020 (2001), hep--ph/0109058, and
JHEP {\bf 0009}, 001 (2000), hep--ph/0007259;
P. Gambino, M. Gorbahn and U. Haisch, Nucl. Phys. {\bf B673}, 238 (2003),
hep--ph/0306079.
%
\bibitem{gambino} 
G. Degrassi, P. Gambino and G.F. Giudice, JHEP {\bf 0012}, 009 (2000),
hep-ph/0009337.  
%
\bibitem{leszek}
K. Okumura and L. Roszkowski, Phys. Rev. Lett. {\bf 92}, 161801 (2004), 
hep--ph/0208101.
%
\bibitem{bsggl}
See e.g. F. Borzumati, C. Greub, T. Hurth and D. Wyler, Phys. Rev. {\bf D62},
075005 (2000), hep--ph/9911245, and references therein.
%
\bibitem{susytalk}
M. Drees, talk held at {\it SUSY2004}, Tsukuba, Japan, June 2004,
hep--ph/0410113.
%
\bibitem{d0top}
D0 Collab. V.M. Abazov et al., Nature {\bf 429}, 638 (2004), 
hep--ex/0406031.
%
\bibitem{pauss}
L.S. Stark P. Hafliger, A. Biland and F. Pauss, JHEP {\bf 0508}, 059 (2005),
hep--ph/0502197.
%
\bibitem{hpole}
H. Baer, A. Belyaev, T. Krupovnickas and X. Tata, JHEP {\bf 0402}, 007 (2004), 
hep--ph/0311351;
H. Baer, T. Krupovnickas and X. Tata, JHEP {\bf 0406}, 061 (2004),
hep--ph/0405058;
A. Djouadi, M. Drees and J.L. Kneur, Phys. Lett. {\bf B624}, 60 (2005),
hep--ph/0504090.
%
\bibitem{dn3}
M. Drees and M.M. Nojiri, Phys. Rev. {\bf D47}, 376 (1993);
T. Falk, R. Madden, K.A. Olive and M. Srednicki, Phys. Lett. {\bf B318}, 354
(1993), hep--ph/9308324.
%
\bibitem{stop-co}
C. Boehm, A. Djouadi and M. Drees, Phys. Rev. {\bf D62}, 035012 (2000), 
hep-ph/9911496; 
R. Arnowitt, B. Dutta and Y. Santoso, Nucl. Phys. {\bf B606}, 59 (2001); 
J.R. Ellis, K.A. Olive and Y. Santoso, Astropart. Phys. {\bf 18}, 395 (2003). 
%
\bibitem{book}
See e.g. M. Drees, R.M. Godbole and P. Roy, {\it Theory and Phenomenology of
  Sparticles}, World Scientific (Singapore, 2004).
%
\bibitem{ell}
U. Ellwanger, Phys. Lett. {\bf B141}, 435 (1984).
%
\bibitem{abrev}
For a recent review, see A. Djouadi, hep--ph/0503173.
%
\bibitem{dn2}
M. Drees and M.M. Nojiri, Phys. Rev. {\bf D45}, 2482 (1992).
%
\bibitem{unit}
K. Griest and M. Kamionkowski, Phys. Rev. Lett. {\bf 64}, 615 (1990).
%
\bibitem{martinvaughn}
S.P. Martin and M.T. Vaughn, Phys. Lett. {\bf B318}, 331 (1993),
hep-ph/9308222. 
%
\bibitem{dn4}
M. Drees and M.M. Nojiri, Phys. Rev. {\bf D48}, 3483 (1993), hep--ph/9307208;
A. Djouadi and M. Drees, Phys. Lett. {\bf B484}, 183 (2000), hep--ph/0004205.
%
\bibitem{sig_foc}
See e.g. J.R. Ellis, J.L. Feng, A. Ferstl, K.T. Matchev and K.A. Olive, 
Eur. Phys. J. {\bf C24}, 311 (2002), astro--ph/0110225.
%
\bibitem{cdms}
CDMS Collab., D.S. Akerib et al., astro--ph/0509259.
%
\bibitem{lhc}
See e.g. H. Baer, C. Balazs, A. Belyaev, T. Krupovnickas and X. Tata, 
JHEP {\bf 0306}, 054 (2003), hep--ph/0304303.
%
\bibitem{tevhiggs}
See e.g. Tevatron Higgs Working Group Collab. M. Carena et al., 
hep--ph/0010338.
%
%
\end{thebibliography}
\end{document}